\documentclass[a4paper,12pt,nosumlimits]{amsart}
\usepackage{latexsym,amsmath,amssymb,enumerate,bbm,fullpage,amscd,graphicx,color}
\usepackage[english]{babel}

\title[Moduli Spaces of Noncommutative Instantons]{Moduli Spaces of Instantons on \\ [4pt] Toric Noncommutative Manifolds}
\author{Simon Brain, Giovanni Landi and Walter D. van Suijlekom}
\address{Unit\'{e} de Recherche en Mathematiques, Universit\'{e} du Luxembourg (Campus Kirchberg), 6 rue Richard Coudenhove-Kalergi, L-1359 Luxembourg, Grand Duchy of Luxembourg}
\address{Dipartimento di Matematica, Universit\`{a} di Trieste, Via A. Valerio 12/1, 34127 Trieste, Italy and INFN, Sezione di Trieste, Trieste, Italy}
\address{Institute for Mathematics, Astrophysics and Particle Physics, Faculty of Science, Radboud University Nijmegen, Heyendaalseweg 135, 6525 AJ Nijmegen, The Netherlands} 
\email{simon.brain@uni.lu, landi@univ.trieste.it, waltervs@math.ru.nl}

\date{10 Apr 2012}

\numberwithin{equation}{section}

\newtheorem{thm}{Theorem}[section]

\newtheorem{lem}[thm]{Lemma}
\newtheorem{prop}[thm]{Proposition}
\newtheorem{rem}[thm]{Remark}
\theoremstyle{definition}
\newtheorem{defn}[thm]{Definition}

\newcommand{\half}{\tfrac{1}{2}}
\newcommand{\Cinf}{C^\infty}
\newcommand{\CMt}{C^\infty(M_\theta)}

\DeclareMathOperator{\ad}{ad}

\renewcommand{\L}{\mathrm{L}}
\newcommand{\Om}{\Omega}

\newcommand{\la}{\langle}
\newcommand{\ra}{\rangle}

\newcommand{\n}{\nabla}
\newcommand{\nom}{\nabla_\omega}
\newcommand{\ep}{\epsilon}
\newcommand{\M}{\textup{M}}
\DeclareMathOperator{\End}{End}
\newcommand{\id}{\textup{id}}
\newcommand{\D}{\textup{d}}
\newcommand{\p}{\partial}
\newcommand{\ii}{\mathrm{i}}
\newcommand{\iGamma}{\Gamma^\infty}
\newcommand{\htimes}{\,\widehat\otimes\,}
\newcommand{\thtimes}{\,\widehat\otimes_\theta\,}

\DeclareMathOperator{\Exp}{Exp}
\DeclareMathOperator{\Ker}{Ker}

\DeclareMathOperator{\im}{Im}
\DeclareMathOperator{\ind}{Index}
\def\cD{\mathcal{D}}
\def\cS{\mathcal{S}}
\def\dR{\mathrm{dR}}

\newcommand{\pp}{{\sf p}}

\newcommand{\sfu}{{\sf u}}

\newcommand{\A}{\mathcal{A}}
\newcommand{\cB}{\mathcal{B}}
\newcommand{\cC}{\mathcal{C}}
\newcommand{\E}{\mathcal{E}}
\newcommand{\cG}{\mathcal{G}}
\newcommand{\h}{\mathcal{H}}
\newcommand{\mM}{\mathcal{M}}

\newcommand{\U}{\textup{U}}
\newcommand{\SU}{\textup{SU}}
\newcommand{\Spin}{\textup{Spin}}
\newcommand{\SO}{\textup{SO}}
\newcommand{\CO}{\mathcal{O}}

\newcommand{\CC}{\mathbb{C}}

\newcommand{\ZZ}{\mathbb{Z}}
\newcommand{\C}{\mathbb{C}}
\newcommand{\TT}{\mathbb{T}}
\newcommand{\RR}{\mathbb{R}}
\newcommand{\R}{\mathbb{R}}

\newcommand{\bz}{{}^{{\scriptscriptstyle(0)}}}

\newcommand{\bo}{{}^{{\scriptscriptstyle(1)}}}

\def\rbiprod{{\cdot\kern-.31em\tr\!\!\!<}}
\def\lbiprod{{>\!\!\!\triangleleft\kern-.33em\cdot}}

\newcommand{\tr}{\triangleright}
\newcommand{\tl}{\triangleleft}

\newcommand{\intdix}{\int\!\!\!\!\!\!\! - ~}

\begin{document}
\begin{abstract}
We study analytic aspects of $\U(n)$ gauge theory over a toric noncommutative manifold 
$M_\theta$. We analyse moduli spaces of solutions to the self-dual Yang-Mills equations on $\U(2)$ vector bundles over four-manifolds $M_\theta$, showing that each such moduli space is either empty or a smooth Hausdorff manifold whose dimension we explicitly compute. In the special case of the four-sphere $S^4_\theta$ we find that the moduli space of $\U(2)$ instantons with fixed second Chern number $k$ is a smooth manifold of dimension $8k-3$.
\end{abstract}

\maketitle

\tableofcontents

\linespread{1.1} 
\parskip 1ex

\subsubsection*{Acknowledgments} SJB was supported by the NWO grant 040.11.163 and by FNR Luxembourg project 894130. He thanks IMAPP, Radboud University Nijmegen, for the kind hospitality during a very enjoyable visit during the period June-December 2010.  \\ GL was partially supported by the Italian
Project ``Cofin08 -- Noncommutative Geometry, Quantum Groups and
Applications''.

\section{Introduction}
In this paper we present a major step forward in our understanding of the differential geometry of toric noncommutative manifolds. Through a series of insightful constructions which generalise a wide range of functional analytic techniques to the setting of noncommutative geometry, we obtain an explicit description of the moduli space of gauge equivalence classes of Hermitian connections on a given toric noncommutative manifold $M_\theta$. In the special case where $M_\theta$ is four-dimensional, we give a detailed analysis of the manifold structure of the moduli space of instanton gauge fields.

Whilst our `noncommutative' moduli space construction is very much modeled upon that of \cite{AHS78} ({\em cf}. also \cite{law}) for classical manifolds, it is not simply a matter of repeating the arguments given there. The vast majority of the needed analytic techniques do not carry over directly from the classical to the noncommutative setting and {\em a priori} we cannot rely upon them. In the present paper we perform a thorough dissection of the smooth structure of toric noncommutative manifolds $M_\theta$, in order to ascertain precisely how the analytic properties of gauge theories thereon are related to those of their classical counterparts.

For example, the methods of \cite{AHS78} rely heavily on properties of elliptic differential operators on compact manifolds, whereas on a noncommutative manifold there is currently no general notion of elliptic theory. We are therefore led to the development of a necessarily more refined analysis of the various (unbounded) linear operators appearing in our construction. Although it is entirely possible that the concrete methods of \cite{ctt,ct} for elliptic operators on the noncommutative torus $\TT^2_\theta$ might admit a generalisation to arbitrary toric noncommutative manifolds $M_\theta$,  herein we derive the required properties of our linear `differential' operators by probing the very delicate and subtle relationships between gauge theories on classical and noncommutative spaces.

As in previous works \cite{bl:mod,brvs}, the crucial tool in the present paper will be a functorial deformation procedure to derive the noncommutative geometry of $M_\theta$ from the classical geometry of $M$ in a systematic way, by deforming along an action of the $N$-torus $\TT^N$. This `quantisation functor' constitutes the foundation upon which our construction is built, in the sense that it explains very precisely which aspects of the classical geometry are preserved by the deformation. Using this framework, one quickly finds that all torus-equivariant features such as the spin and metric structures are canonically deformed. However, we stress that the construction of instantons on a toric noncommutative four-manifold $M_\theta$ is much more subtle and not simply obtained through a functorial quantisation of classical instantons. 
The reason for this is simply that not every classical instanton is given as a torus-equivariant connection, and so cannot be deformed directly. 

The paper is organised as follows. In \S\ref{se:prelims} we set the stage by introducing the relevant category theory we shall need in order to describe our functorial deformation procedure. In \S\ref{se:ncmflds} we apply this deformation theory to explicitly derive the differential geometry of the toric noncommutative manifold $M_\theta$ obtained by functorial quantisation of a given classical manifold $M$ with a torus action. As already mentioned, the functorial nature of the deformation means that it simultaneously deforms all torus-equivariant geometric structures, including vector bundles, principal bundles and associated spin structure. 

In \S\ref{se:ins} we elaborate upon the notions of $\U(n)$ gauge theory on a toric noncommutative manifold, including the gauge group of a noncommutative vector bundle, the notion of a connection on such a bundle and what it means for connections to be gauge equivalent. Finally, we discuss Dirac operators on $M_\theta$ with coefficients in a noncommutative vector bundle and derive a Weitzenb\"ock formula for toric noncommutative manifolds, relating the square of a Dirac operator with coefficients to the curvature of the underlying connection. 

In \S\ref{se:eq-classes} there are the functional analytic details of the moduli space construction. In analogy with \cite{law}, we introduce Sobolev norms on the affine space of connections on a given vector bundle and equip the gauge group with the structure of a Banach-Lie group. This eventually enables us to endow with a Banach manifold structure the (infinite-dimensional) space of gauge equivalence classes of Hermitian connections on a given noncommutative vector bundle over $M_\theta$. In lieu of a suitable reference, we present the analytical details of this construction in full detail, making our exposition somewhat self-contained.

In \S\ref{se:mod-sp} we specialise to $M_\theta$ being a four-dimensional noncommutative manifold and study in detail the manifold structure of the moduli space of instantons. An index argument expresses the dimension of the moduli space in terms of Chern classes of the underlying vector bundles, in parallel with the classical analysis \cite{AHS78} although now in noncommutative parlance. In the special example of the toric noncommutative four-sphere $S^4_\theta$, we find that the moduli space of instantons on a vector bundle with fixed topological charge $k\in\ZZ$ has dimension $8k-3$, in agreement with the value suggested by the results of \cite{lvs:pfns} and \cite{brvs}.

\section{Categorical Preliminaries}
\label{se:prelims} 

We will obtain noncommutative manifolds from classical ones through a deformation procedure which is categorical; the deformation itself will have the form of a functor sending the classical geometry to the noncommutative geometry. We begin in this section by introducing the various categories and methods that we need.

\subsection{Objects with a torus action}
We study the deformation of manifolds 
along an isometric action of a real $N$-torus $\TT^N$ with $N\geq 2$. Thus the first category we need is the  collection of spaces (with some structure) which carry such an action of $\TT^N$.

Let $V$ be a nuclear Fr\'{e}chet space whose topology is determined by a countable family $\|\cdot\|_j$ of semi-norms.
Suppose that $V$ is equipped with a smooth 
action $\alpha:\TT^N\times V\to V$ of the torus $\TT^N$.
This action is taken to be isometric with respect to the family $\|\cdot\|_j$. Given a pair of such nuclear Fr\'{e}chet $\TT^N$-modules $(V,\alpha)$ and $(W,\beta)$, a continuous linear transformation $T:V\to W$ is said to be {\em $\TT^N$-equivariant} if there is a commutative diagram
$$
\begin{CD} 
{V} @>T>> {W}
\\ @VV{\alpha} V @VV {\beta} V \\ {V} @>T>> {W} 
\end{CD} \quad ,
$$
that is to say, $T$ is an intertwiner for the actions $\alpha$ and $\beta$.
\begin{defn}
We write $\mathcal{V}_N$ for the category whose objects are pairs $(V, \alpha)$ as above and whose morphisms are continuous $\TT^N$-equivariant linear transformations. When there is no possibility of confusion, we shall omit the subscript $N$ and write $\mathcal{V}=\mathcal{V}_N$.
\end{defn}

Very important for us is the fact that the category $\mathcal{V}$ is {\em monoidal}. Indeed, given objects $(V,\alpha)$ and $(W,\beta)$ in $\mathcal{V}$, the algebraic tensor product $V\otimes W$ carries the diagonal action $\alpha\otimes\beta$ of the torus $\TT^N$. One equips $V\otimes W$ with the family of semi-norms\footnote{The fact that the families of semi-norms on $V$ and $W$ may always be assumed to be increasing \cite{mr:def} makes it possible to define the semi-norms on the tensor product in this way.}
\begin{equation}\label{cross}
\|x\|_j:=\textup{inf}\,\left\{\sum_r \|v^r\|_j \|w^r\|_j~:~x=\sum_r v^r\otimes w^r\right\},
\qquad x\in V\otimes W .
\end{equation}
The completion $V\htimes W$ of $V\otimes W$ in the topology defined by these semi-norms is again a Fr\'{e}chet space (the nuclearity assumption assures that this is unambiguously defined). The diagonal action $\alpha\otimes\beta$ extends to an action $\alpha\htimes\beta$ on $V\htimes W$ and the resulting pair  $(V\htimes W,\alpha\htimes\beta)$ is an object in the category $\mathcal{V}$.
Moreover, the category $\mathcal{V}$ is {\em braided}: for a pair of objects $(V,\alpha)$ and $(W,\beta)$ there is a continuous torus-equivariant isomorphism
\begin{equation}\label{braid}
\Psi^0_{V,W}:V\otimes W\to W\otimes V,\qquad \Psi^0_{V,W}(v\otimes w)=w\otimes v ,
\end{equation}
for each $v\in V$ and $w\in W$, which extends to an isomorphism $\Psi^0_{V,W}:V\htimes W\to W\htimes V$. 

In fact, we shall also need to deal with actions of a covering torus $c:\widetilde \TT^N\to
\TT^N$. We write $\widetilde{\mathcal{V}}=\widetilde{\mathcal{V}}_N$ for the category whose objects are nuclear Fr\'{e}chet $\widetilde\TT^N$-modules $(V,\alpha)$ and whose morphisms are Fr\'{e}chet-continuous $\widetilde\TT^N$-equivariant linear maps. The following lemma tells how to relate the actions of these two tori on a given vector space.

\begin{lem}\label{le:catinc} Let $c:\widetilde \TT^N\to
\TT^N$ be a covering of the torus $\mathbb{T}^N$. Then there is a
full embedding of monoidal categories
$\mathcal{V}\hookrightarrow \widetilde{\mathcal{V}}$.\end{lem}

\proof The functor $\mathcal{V}\hookrightarrow \widetilde{\mathcal{V}}$ is
the one which assigns to each $\TT^N$-module $(V,\alpha)$ the  $\widetilde\TT^N$-module $(V,c^*\alpha)$, which is the same Fr\'{e}chet space $V$ equipped with the pull-back action along the covering map $c:\widetilde \TT^N\to
\TT^N$. A little thought shows that this functor is well-defined and that it is a full monoidal embedding.\endproof

Thus, if we start with the category $\mathcal{V}$, when  
encountering a covering torus $\widetilde\TT^N$ of $\TT^N$ we can deal with it by passing to
$\widetilde{\mathcal{V}}$ via the above embedding. Henceforth we shall always tacitly assume that this technical point has been
dealt with and pay it no further attention.

In order to describe the deformation procedure, first of all we need to introduce some notation. With $\RR^N$ the Lie algebra of $\TT^N$, we may lift the torus action $\alpha$ to a periodic action of $\RR^N$ with kernel the integer lattice $\ZZ^N$ in $\RR^N$, so that $\TT^N\simeq \RR^N/\ZZ^N$.  

Let $(A,\alpha)$ be an object in the category $\mathcal{V}$.
With $G=\RR^N\times \RR^N$, consider the space $C^\infty(G,A)$ of smooth bounded functions from $G$ to $A$. Given a choice of basis $(x_1,\dots, x_N)$ for $\RR^N$ we let $\p_{k}$ denote the operation of partial differentiation on $C^\infty(G,A)$ in the direction of $x_k$, for $k=1,\ldots,2N$. For a multi-index 
$\mu=(\mu_1,\ldots,\mu_{2N})$ we write $\p^{(\mu)}$ for the corresponding higher partial differentiation. 
Then $C^\infty(G,A)$ is a Fr\'{e}chet space with respect to the family of semi-norms
\begin{equation}\label{t-norm}
\|F\|_{j,k}:=\textup{sup}_{i\leq j}  \sum_{|\mu|\leq k} \ \frac{1}{\mu!}\ \textup{sup}_{u\in G} \|  \p^{(\mu)}  F(u)\|_i,
\end{equation}
where $\mu!:=\mu_1!\ldots\mu_N!$ is a normalisation factor, and the action of $G$ on $C^\infty(G,A)$ by translation is both smooth and isometric. Take $\RR^N$ to be equipped with an inner product, which we denote as a dot product, and choose a Haar measure on $\RR^N$, and hence on $G$, such that the Fourier transform is unitary. Finally, let $e:\RR\to\CC$ denote the complex-valued function $e(t):=\exp (\pi\ii \, t)$. Amongst other things, it is shown in \cite[Ch.~1]{mr:def} that for any function $F\in C^\infty(G,A)$ the following integral is well-defined, 
\begin{equation}\label{gen-int}
\int_{G} F(x,y) \, e(x\cdot y)\ \D x \D y \, ,
\end{equation}
and that, for all $j$ and all $F$, there exists an index $k$ and a constant $c_k$ such that
\begin{equation}\label{est}
\left\|\int_{G} F(x,y) \, e(x\cdot y)\ \D x \D y \right\|_j \leq c_k\|F\|_{j,2k} \, .
\end{equation}
Such estimates are the cornerstone of the deformation theory of \cite{mr:def} and they will also prove crucial in what follows here.

\subsection{The deformation functor}\label{se:quant} 
Having established a category of spaces comprising the relevant classical geometry, we now describe the procedure which will give rise to the corresponding noncommutative geometry.

Let $\theta:\RR^N\to \RR^N$ be a skew-symmetric linear transformation. Starting from the monoidal category 
$\mathcal{V}$ we define a new monoidal category $\mathcal{V}_\theta$ as follows. The objects and morphisms are taken to be the same as
those of $\mathcal{V}$, so the category $\mathcal{V}_\theta$ again consists of nuclear Fr\'{e}chet $\TT^N$-modules and continuous torus-equivariant linear transformations. 
In this way, there is an obvious functor 
$$
\L_\theta:\mathcal{V}\to \mathcal{V}_\theta
$$ 
which is just the identity functor, {\em i.e.} it leaves objects and morphisms unchanged. 

The crucial point, and more interesting, is the way in which we turn $\L_\theta$ into a monoidal functor, leading to a monoidal structure $\thtimes$ for $\mathcal{V}_\theta$. Indeed, we define the tensor product $\L_\theta (V)\thtimes \L_\theta(W)$ to be equal to  $\L_\theta(V\htimes W)$ as a Fr\'{e}chet $\TT^N$-module (so, in particular, the torus action and semi-norms are just the same as they were in the category $\mathcal{V}$). However, as vector spaces, we do not take the trivial identification but instead seek to define a map
$$
c^{}_{V,W}:\L_\theta (V)\thtimes \L_\theta(W)\to \L_\theta(V\htimes W), 
$$
given on indecomposable tensors by 
\begin{equation}\label{th-iso}
c_{V,W}(v\otimes_\theta w):=\int_{G}\left(\alpha_{\theta x}(v)\otimes\beta_y(w)\right) \, 
e(x\cdot y)\ \D x\D y, \quad \textup{for all}~~ v\in (V,\alpha), ~w\in (W,\beta),
\end{equation}
with the symbol $\otimes$ in the integral denoting the usual tensor product (and $G=\RR^N\times \RR^N$ and the Haar measure as before). 
The integral is easily seen to be well-defined due to the expression \eqref{gen-int} being well-defined, taking 
$A=V\htimes W$ and $F(x,y)=\alpha_{\theta x}(v)\otimes\beta_y(w)$ for fixed $v \in V$ and $w \in W$, clearly a smooth bounded function in its two arguments $x,y$. 
We next show that this map is continuous and extends to the completion $\L_\theta(V)\thtimes\L_\theta(W)$. Although this was already pointed out in \cite{hm}, we include the details here since the result will play such an important part in the following.

\begin{lem}
For all $j$ there exists an index $m$ and a constant $d_m$ such that
$$
\|c_{V,W}(z)\|_j\leq d_m \|z\|_m \qquad \textup{for all}~~ z\in\L_\theta (V)\otimes_\theta \L_\theta(W), 
$$
where $\|\cdot\|_j$ are the semi-norms on $\L_\theta (V)\thtimes \L_\theta(W)\simeq V\htimes W$ defined in Eq.~\eqref{cross}.
\end{lem}

\proof From the inequality \eqref{est}, again with $A=V\htimes W$ and $F(x,y)=\alpha_{\theta x}(v)\otimes\beta_y(w)$, we know that for all $j$ there exists a $k$ and a constant $c_k$ such that
\begin{equation}\label{ineq-1}
\|c_{V,W}(v\otimes_\theta w)\|_j\leq c_k \|F\|_{j,2k} 
\end{equation}
on indecomposable tensors. By differentiating the action (by translation) of $G$ on the function $F$, for any $l=1,\ldots,N$ we find that 
\begin{align*}
\p_{l,x}F (x,y)&=\textup{lim}_{h\to 0} ~h^{-1}\left( \alpha_{\theta(x+h X_l)}(v)-\alpha_{\theta x}(v)\right)\otimes \alpha_y(w) \\
&=\textup{lim}_{h\to 0} ~  \alpha_{\theta x} \left( h^{-1}(\alpha_{h\theta X_l}(v)-v)\right)\otimes\alpha_y(w) \\
&=\alpha_{\theta x}\left(\sum_m \theta_{lm} \textup{lim}_{h\to 0} ~h^{-1}(\alpha_{hX_m}(v)-v)\right)\otimes\alpha_y(w) \\
&=\sum_m \theta_{lm}\alpha_{\theta x}(\p_m v)\otimes \alpha_y(w) . 
\end{align*}
Here we have used the standard properties of the group action, together with its continuity to move the limit inside the sum. Similarly, one finds
$$
\p_{l,y}F=\alpha_{\theta x}(v)\otimes \alpha_y(\p_l w) .
$$
Since $\alpha$ is isometric, it follows that for any $i$ there is a constant $c$ for which 
$$ 
\|\p_{l,x}F\|_i\leq c~ \textup{sup}_m \|\p_m(v)\|_i\|w\|_i  \qquad \textup{and} \qquad  
\|\p_{l,y}F\|_i\leq\|v\|_i\|\p_l(w)\|_i. 
$$
Estimates for the higher derivatives of $F$ are obtained in a similar way. Thus, from the formula \eqref{t-norm} for the norm $\|\cdot\|_{j,2k}$ and using boundedness of the operators $\p^{(\mu)}$ on smooth vectors $v$ and $w$, we deduce that there exists an index $m$ and a constant $c'_m$ such that
\begin{equation}\label{ineq-2}
\|F\|_{j,2k}\leq c'_m\|v\|_m\|w\|_m.
\end{equation}
Combining the inequalities \eqref{ineq-1} and \eqref{ineq-2} and combining together the constants, we deduce  
there exists an index $m$ and a constant $d_m$ such that
$$
\|c_{V,W}(v\otimes_\theta w)\|_j\leq d_m \|v\|_m\|w\|_m.
$$
Now let $z\in \L_\theta (V)\thtimes\L_\theta(W)$ be an element of the algebraic tensor product and choose a representative $z=\sum_r v^r\otimes_\theta w^r$. Just as above we find that there exists an $m$ and a constant $d_m$ such that
$$
\|c_{V,W}(z)\|_j\leq d_m \sum_r\|v^r\|_m\|w^r\|_m.
$$
Taking the infimum on the right hand side over all such representatives of $z$ gives the inequality stated in the lemma.
\endproof
\noindent
It follows that the map $c_{V,W}$ is continuous and so we may indeed extend it to a map 
\begin{equation}\label{cons}
c_{V,W}:\L_\theta(V)\thtimes\L_\theta(W)\to\L_\theta(V\htimes W),
\end{equation}
as was our aim. It is easily seen that $c_{V,W}$ is invertible with continuous inverse defined by the transformation $\L_{\theta^{\textup{t}}}$, with $\theta^{\textup{t}}$ denoting the transpose matrix, and hence an isomorphism of Fr\'{e}chet $\TT^N$-modules \cite{mr:def,hm}. 
The next step uses the map $c_{V,W}$ and its inverse to define a natural braiding on $\mathcal{V}_\theta$, yielding the following result.

\begin{prop}
The functor $\L_\theta:\mathcal{V}\to \mathcal{V}_\theta$ is an isomorphism of braided monoidal categories.
\end{prop}

\proof 
By definition the functor $\L_\theta$ is an isomorphism of categories. 
With $\Psi^0$ the braiding on $\mathcal{V}$ defined in Eq.~\eqref{braid},  
we equip the category $\mathcal{V}_\theta$ with the braiding defined by
\begin{equation}\label{th-braid}
\Psi_{V,W}:V\thtimes W\to W\thtimes V,\quad \Psi_{V,W}:=c_{V,W}^{-1}\circ\Psi^0_{V,W}\circ c^{}_{V,W} ,
\end{equation}
for each pair of objects $V,W$ in $\mathcal{V}_\theta$. 
Continuity of $\Psi$ follows immediately from its being the composition of continuous maps. It is then straightforward to check that $\L_\theta$ is an intertwiner for the braidings \eqref{braid} and \eqref{th-braid} on the categories $\mathcal{V}$ and $\mathcal{V}_\theta$, respectively.\endproof

These structures are particularly
useful when examining what happens to algebras in the category
$\mathcal{V}$ under the functor $\L_\theta$. Let $\A$ be an algebra in
$\mathcal{V}$ so that, in particular, the product map $m:\A\htimes \A\to \A$ is continuous and torus-equivariant. 
Via $\L_\theta$ the product $m$
becomes a map $\L_\theta(\A\htimes \A) \to \L_\theta(\A)$. Composing it with
$c^{}_{\A,\A}$ yields a new product map
\begin{equation}\label{twcopr}
m_\theta:\L_\theta(\A)\thtimes \L_\theta(\A)\to \L_\theta(\A), \qquad m_\theta:=m\circ c^{}_{\A,\A}.
\end{equation}
Since the product $m_\theta$ is a composition of continuous torus-equivariant maps, it also shares these properties, making $\L_\theta(\A)$ into a Fr\'{e}chet algebra in the category $\mathcal{V}_\theta$. In this way we think of the functor $\L_\theta$ as a `quantisation
functor', since it gives a way of simultaneously deforming all algebras in
$\mathcal{V}$ into new algebras with deformed products which are nevertheless torus-equivariant and continuous.

More generally, let $\A$ be an algebra in $\mathcal{V}$ and let $\E$ be 
an $\A$-bimodule in $\mathcal{V}$ (or simply a left module or a right module). 
This means that $\E$ is itself a nuclear Fr\'{e}chet space such that the left and/or right module structures
$$
\tr:\A\htimes \E\to \E,\qquad \tl:\E\htimes \A\to \E,
$$
are continuous and torus-equivariant. Then, under the deformation functor $\L_\theta$, one automatically finds that $\L_\theta(\E)$ is a (bi)module over the algebra $\L_\theta(\A)$ in the category $\mathcal{V}_\theta$ when equipped with the deformed left and/or right module structures
$$
\tr_\theta:=\tr\circ c^{}_{\A,\E},\qquad \tl_\theta:=\tl\circ c^{}_{\E,\A},
$$
in the sense that these maps are automatically torus-equivariant and continuous.

Given the general categorical framework described above, in `practical situations' one may use 
a simplified version of the deformation procedure, which we now illustrate. 
Recall that the Pontryagin dual of the torus group $\TT^N$ is the additive group $\ZZ^N$. Given an object $(V,\alpha)$ of the category $\mathcal{V}$, every element $v\in V$ has a unique series decomposition
$$
v=\sum_{r\in\ZZ^N}\, v_r
$$
which is rapidly convergent in the Fr\'{e}chet topology on $V$. Here, for $r\in\ZZ^N$, each term $v_r$ is 
{\em homogeneous of degree $r\in \ZZ^N$}, that is to say it is such that 
$$
\alpha(t)(v_r)=e(r\cdot t)\, v_r, \qquad \textup{for} \quad t\in \TT^N . 
$$
We write $V_r$ for the vector subspace of $V$ consisting of homogeneous elements of degree $r\in \ZZ^N$. 
We thus have a direct sum decomposition
$
V=\oplus_{r\in\ZZ^N}\, V_r
$
and so it follows that equipping the Fr\'{e}chet space $V$ with a smooth isometric action of $\TT^N$ is equivalent to giving a $\ZZ^N$-grading on $V$. In these terms, on homogeneous elements $v_r\in V$ and $w_s\in W$ the 
map \eqref{th-iso} has the simple form
$$
c_{V,W}(v_r\otimes_\theta w_s)=\chi(r,s)\, v_r\otimes w_s,
$$
where $\chi$ is the bi-character on $\ZZ^N\times \ZZ^N$ defined by $\chi(r,s)=e (r\cdot\theta s)$.

\subsection{A noncommutative Hopf fibration}\label{se:qhopf} Here we
provide an instance of the deformation functor in operation: we use
it to deform the $\SU(2)$ Hopf fibration $S^7\to S^4$, an example
which has proved central in the study of noncommutative instantons
and will provide us with our main example at the end of the paper.

The algebra $C^\infty(S^4)$ of functions on the sphere $S^4$ is the
commutative unital smooth $*$-algebra generated by the coordinate functions
$x_1$, $x_2$, their conjugates $x_1^*$, $x_2^*$ and the
self-conjugate element $x_0=x_0^*$, subject to the sphere relation
\begin{equation}\label{foursph}
x_1^*x_1+x_2^*x_2+x_0^2=1.
\end{equation} 
The algebra $\Cinf(S^4)$ carries
a smooth action of the two-torus $\TT^2$ given on generators by
\begin{equation}\label{coactfour}
\TT^2\times \Cinf(S^4)\to \Cinf(S^4),\qquad x_1 \mapsto e^{2\pi\ii t_1} x_1,
\quad x_2\mapsto e^{2\pi\ii t_2} x_2,\quad x_0\mapsto x_0,
\end{equation}
where $(e^{2\pi\ii t_1},e^{2\pi\ii t_2})\in \TT^2$. This action makes $C^\infty(S^4)$ into an algebra in the category 
$\mathcal{V}_2$.

Similarly, the algebra $\Cinf(S^7)$ of smooth functions on the sphere $S^7$ is
the commutative unital smooth $*$-algebra generated by the coordinate
functions $z_j$, $j=1,\ldots,4$, together with their conjugates
$z_j^*$, $j=1,\ldots,4$, subject to the sphere relation
\begin{equation}\label{sevsph}
z_1^*z_1+z_2^*z_2+z_3^*z_3+z_4^*z_4=1.
\end{equation}
This time we take a covering two-torus $\widetilde{\TT}^2$ (see below for the rationale for this choice) with coordinates $(e^{2\pi\ii \tilde t_1},e^{2\pi\ii \tilde t_2})\in\widetilde{\TT}^2$ and define an action $\widetilde{\TT}^2\times \Cinf(S^7)\to \Cinf(S^7)$ by
\begin{eqnarray}\label{coactsph}
(z_1,z_2,z_3,z_4)\mapsto (e^{2\pi\ii \tilde t_1}z_1,e^{-2\pi\ii\tilde t_1}z_2,e^{2\pi\ii\tilde t_2}z_3,e^{-2\pi\ii \tilde t_2}z_4).
\end{eqnarray}
This action makes $\Cinf(S^7)$ into an algebra in the category $\widetilde{\mathcal{V}}_2$.

If one arranges the generators of the algebra $\Cinf(S^7)$ into the matrix
\begin{equation}\label{u}
\sfu:=\begin{pmatrix} z_1 & -z_2^* \\ z_2 & z_1^* \\ z_3 & -z^*_4 \\ z_4 &z_3^*\end{pmatrix} ,
\end{equation}
there is a right action of the classical group $G=\SU(2)$ on $\Cinf(S^7)$ given by
\begin{equation}\label{gpact}
\Phi:\Cinf(S^7)\times G\to \Cinf(S^7),\qquad \Phi_g(\sfu)=\sfu
g, 
\end{equation}
with $g\in\SU(2)$ in its fundamental $2\times 2$ matrix representation. This action is well-defined since it preserves the sphere relation
\eqref{sevsph}. The invariant subalgebra under this action is found
to be isomorphic to $\Cinf(S^4)$ via the identification
\begin{equation}\label{alginc}x_1=2(z_1z^*_3 + z_2^*z_4), \quad x_2=2(z_2z_3^* -
z_1^*z_4), \quad x_0= z_1z^*_1 + z_2z^*_2 - z_3z_3^* -
z_4z_4^* .
\end{equation} 
It follows that Eqs.~\eqref{alginc} define an inclusion of algebras
$\Cinf(S^4)\hookrightarrow\Cinf(S^7)$, yielding a `dual' description of the standard Hopf
fibration $S^7\to S^4$ with $\SU(2)$ as structure group.

\begin{rem}\textup{There is an obvious covering map of abelian groups given by
$$
c:\widetilde{\TT}^2\rightarrow\TT^2,\qquad
(e^{2\pi\ii t_1},e^{2\pi\ii t_2})= (e^{2\pi\ii (\tilde t_2+\tilde t_1)},e^{2\pi\ii (\tilde t_2-\tilde t_1)}).
$$
This covering is compatible with the actions \eqref{coactfour} and
\eqref{coactsph}, giving an illustration of Lemma~\ref{le:catinc} and 
we may as well assume we are working exclusively in the
category $\widetilde{\mathcal{V}}_2$.}
\end{rem}

We apply the deformation theory of \S\ref{se:quant} by choosing
a $2\times 2$ real skew-symmetric matrix
$$
\Theta=\begin{pmatrix}0&\theta\\-\theta&0\end{pmatrix} ,  \qquad \textup{with} \quad \theta \in \R,
$$
and defining a bi-character $\chi:\ZZ^2\times \ZZ^2\to \C$ by
$$
\chi(r,s)=\textrm{exp}(\mathrm{i}\pi \, r\cdot\Theta s).
$$
Denoting the generators of $\ZZ^2$ by
$$
(r_1,r_2,r_3,r_4):=\big( (1,0),(-1,0),(0,1),(0,-1) \big) ,
$$
it is clear from the formula \eqref{coactsph} that the generators $z_1,\ldots z_4$ of $\Cinf(S^7)$ have homogeneous degree $r_1,\ldots,r_4$, respectively. The product on $\Cinf(S^7)$ is deformed by
applying the deformation functor and using the formula
\eqref{twcopr} to obtain a new product
$$
z_j\cdot_\chi z_l=\chi(r_j,r_l)z_jz_l,\qquad
z_j\cdot_\chi z_l^*=\chi(r_j,r_l^*)z_jz_l^* \, , \qquad \textup{for} \quad j,l = 1, \ldots,4 .
$$
Introducing the deformation parameter
$\eta_{jl}:=\chi^{-2}(r_j,r_l)$ given explicitly by
\begin{equation}\label{eta}
(\eta_{jl})=\begin{pmatrix}1&1&\mu&\bar\mu\\1&1&\bar\mu&\mu\\\bar\mu&\mu&1&1\\\mu&\bar\mu&1&1\end{pmatrix},
\qquad \mu=e^{\ii\pi\theta},
\end{equation}
the deformed algebra relations are computed to be (dropping the
product symbol $\cdot_\chi$)
$$
z_j\, z_l=\eta_{lj}z_l\, z_j,\qquad z_j\, z_l^*=\eta_{jl}\, z_l^*\,z_j, \qquad \textup{for} \quad j,l = 1, \ldots,4.
$$
On the other hand, the torus action preserves the sphere relation \eqref{sevsph} and so the radius element is not deformed.
We denote by $\Cinf(S^7_\theta)$ the smooth unital $*$-algebra generated by
$\{z_j,z_j^*~|~j=1,\ldots,4\}$ modulo the algebra relations above, and with the sphere relation. 

Similarly, the product on the algebra $\Cinf(S^4)$ is twisted into a new product
$$
x_1\cdot_\chi x_2=\chi(r_1+r_4,r_2+r_4)x_1x_2,\qquad
x_1\cdot_\chi x_2^*=\chi(r_1+r_4,r_1+r_3)x_1x_2^*, 
$$
and products with the generator $x_0$ remain undeformed. 
With deformation parameter $\lambda:=\mu^2=e^{\ii 2 \pi\theta}$, the
relations become (again dropping the product symbol $\cdot_\chi$)
$$
x_1x_2 = \lambda x_2x_1,\quad
x_1^*x_2^*=\lambda x_2^*x_1^*, \quad
x_2^*x_1=\lambda x_1x_2^*,\quad
x_2x_1^*=\lambda x_1^*x_2,
$$
with $x_0$ central. Again the radius element is not deformed, with the relation \eqref{foursph} 
unchanged. We denote by $\Cinf(S^4_\theta)$ the smooth unital $*$-algebra
generated by $x_1,x_2,x_0$ and their conjugates modulo these
new algebra relations, together with the sphere relation.

Since the action \eqref{coactsph} of $\widetilde{\TT}^2$ on $\Cinf(S^7)$ commutes
with the $\SU(2)$-action \eqref{gpact}, the deformation of the
spheres $\Cinf(S^7)$ and $\Cinf(S^4)$ preserves this action and hence
there is an algebra inclusion
$\Cinf(S^4_\theta)\hookrightarrow\Cinf(S^7_\theta)$, once again determined by
Eqs.~\eqref{alginc} on generators. For later use we observe that the sphere
relation \eqref{sevsph} in $\Cinf(S^7_\theta)$ implies that $\sfu^*\sfu=1$, whence the matrix-valued function
\begin{equation}\label{qbas}
\pp:=\sfu\sfu^*=\tfrac{1}{2}\begin{pmatrix} 1+x_0 & 0 & x_1 & -\bar
\mu\, x_2^* \\ 0 & 1+x_0 & x_2 & \mu\, x_1^* \\ x_1^* &
x_2^* & 1-x_0 & 0 \\ -\mu\, x_2 & \bar\mu\, x_1 & 0 & 1-x_0
\end{pmatrix} 
\end{equation}
automatically obeys $\pp^2=\pp=\pp^*$, that is to say $\pp$ is a projection.

We remark that {\em a priori} we could have considered a more
general torus action on the algebras $\Cinf(S^7)$ and $\Cinf(S^4)$ than 
the one given above; but the latter is the most general one which is compatible with the $\SU(2)$-action and so leading to a deformed principal bundle with classical structure group \cite{lvs:pfns}. 

The construction above characterizes the noncommutative four-sphere $S^4_\theta$ in terms of the algebra $\Cinf(S^4_\theta)$ of its smooth functions. 
Fr\'{e}chet algebras of functions on general toric noncommutative manifolds will be the subject of the next section.

\section{Toric Noncommutative Manifolds}\label{se:ncmflds}

In this section we shall apply the general quantisation procedure we outlined in \S\ref{se:quant} to the
function algebra over any compact manifold $M$ carrying an appropriate torus action. 
Being functorial, the quantisation deforms not just the algebra itself but any associated
$\TT^N$-equivariant construction on $M$. 
We use this fact to deform in particular
its differential and metric
structures, together with all equivariant vector and principal bundles over $M$.

\subsection{Torus-equivariant classical geometry}\label{se:toruseq}
Before we come to the deformation procedure, we write all of the necessary geometric
ingredients in the appropriate categorical manner, from which their quantisation will
follow naturally.

Let $(M,g)$ be an $m$-dimensional compact manifold with Riemannian
metric $g$ and assume that $M$ is equipped with a smooth isometric
action $\sigma$ of an $N$-torus $\TT^N$, $N\geq 2$. We also denote
by $\sigma$ the corresponding action of $\TT^N$ by $*$-automorphisms
on the algebra $C^\infty(M)$ of smooth functions on $M$ obtained by
pull-back:
$$
(\sigma_s(f))(x):=f(s^{-1}x), \qquad \text{for}~~  s\in\TT^N,~f\in
C^\infty(M),~x\in M.
$$
The algebra $\Cinf(M)$ comes equipped with a countable family of semi-norms, defined as usual in terms of local partial derivatives, making it into a nuclear Fr\'{e}chet space whose product is continuous with respect to the resulting topology \cite{sha}.
The norm 
\begin{equation}\label{C-norm}
\|\cdot\|: \Cinf(M)\to\C, \qquad \|f\|:=\mathrm{sup}_{x\in M}\, |f(x)| ,
\end{equation}
makes $\Cinf(M)$ a pre-$C^*$-algebra inside the algebra $C(M)$ of continuous functions on $M$.
Finally, there is an integration map for the Riemannian measure of the metric $g$,
\begin{equation}\label{int0}
\int_M:\Cinf(M)\to\C  .
\end{equation}

\begin{lem}\label{le:grad} The algebra $C^\infty(M)$ is an algebra in the category
$\mathcal{V}$.
The norm \eqref{C-norm} and the integration map \eqref{int0} are morphisms in the category.
\end{lem}

\proof As already mentioned, it is a standard fact that $\Cinf(M)$ is a nuclear Fr\'{e}chet algebra. Since the action of $\TT^N$ on $M$ is by smooth isometries, the corresponding action $\sigma$ of $\TT^N$ on $\Cinf(M)$ is smooth and isometric with respect to the 
family of semi-norms, whence $\Cinf(M)$ is an object in the category $\mathcal{V}$.

To see that $C^\infty(M)$ is in the category also an algebra, it is enough to
observe that its product is such that, for each
$r,r'\in \ZZ^N$ and each pair $f_r$, $g_{r'}$ of corresponding homogeneous elements for the action of $\TT^N$, every product $f_r g_{r'}$ belongs to the $(r+r')$-graded subspace of $C^\infty(M)$, so the product is torus-equivariant. 

It is immediate that the norm $\|\cdot\|$ is invariant
for the action of $\TT^N$ on $\Cinf(M)$. Viewing the target
space $\C$ as a trivial $\TT^N$-module, this
invariance property is equivalent to the map
$\|\cdot\|:\Cinf(M)\to \C$ being a morphism in $\mathcal{V}$.
Finally, since the action of $\TT^N$ on $M$ is isometric, it is measure-preserving and so the integration map $f\mapsto \int_M f$ is also torus-invariant, thus a morphism in $\mathcal{V}$. \endproof
 
Let $E$ be a smooth  Hermitian vector bundle over $M$ and write $\iGamma(M,E)$ for the
$C^\infty(M)$-bimodule of smooth sections of $E$. 
The Hermitian structure is determined by a fibre metric, about which we shall say more later on.
With $M$ being a smooth compact manifold, the space $\iGamma(M,E)$ is nuclear Fr\'{e}chet and equipped with commuting
continuous left and right $\Cinf(M)$-module structures \cite{sha}, {\em i.e.} $\iGamma(M,E)$ is a Fr\'{e}chet $\Cinf(M)$-bimodule. 
We suppose the bundle $E$ carries a torus action as well. More precisely, we suppose 
there exists a covering $c:\widetilde\TT^N\to\TT^N$ of the torus
$\mathbb{T}^N$ such that $E$ is equipped with an action
$\tilde\sigma$ of $\widetilde \TT^N$ by bundle automorphisms which
covers the action $\sigma$ of $\TT^N$ on $M$:
\begin{equation}
\label{eq:torus-bundle}
\tilde\sigma_s(\psi f)=\tilde\sigma_s(\psi)\sigma_{c(s)}(f), \qquad \text{for}~~ 
 s\in \widetilde\TT^N  \quad \text{and}~~ 
\psi\in \iGamma(M,E),~f\in C^\infty(M).
\end{equation}
If this is the case, one says that $E$ is a {\em $\tilde\sigma$-equivariant vector bundle}
over $M$.
Given a $\tilde\sigma$-equivariant vector bundle $E$ over $M$, recall that a smooth section $\psi_r \in \iGamma(M,E)$ is said to be {\em homogeneous of degree $r\in \ZZ^N$} if it has the property
$$
\tilde\sigma_s(\psi_r)=e^{2\pi\ii \, r\cdot s}\psi_r \qquad \text{for all}~~ s\in\widetilde\TT^N.
$$

\begin{prop}\label{pr:vecobj} 
Let $E$ be a $\tilde\sigma$-equivariant vector bundle over $M$.
Then $\iGamma(M,E)$ is an object in the
category $\widetilde{\mathcal{V}}$ of Fr\'{e}chet $\widetilde{\TT}^N$-modules. The left and right $C^\infty(M)$-module
structures on $\iGamma(M,E)$ are both morphisms in the
category.\end{prop}

\proof 
As already mentioned, the space $\iGamma(M,E)$ is a nuclear Fr\'{e}chet $\Cinf(M)$-bimodule, {\em i.e.} a nuclear Fr\'{e}chet space 
equipped with commuting continuous left and right $\Cinf(M)$-module structures. Since each section $\psi\in\iGamma(M,E)$ is locally the direct sum of smooth functions, the action of $\widetilde\TT^N$ on $\iGamma(M,E)$ is also smooth and isometric.

Just as was the case for $\Cinf(M)$, every section $\psi\in\iGamma(M,E)$ has the form
\begin{equation}\label{hdp}
\psi=\sum_{r\in\ZZ^N}\psi_r ,
\end{equation}
with $\psi_r$ homogeneous of degree $r\in\ZZ^N$. This defines a $\ZZ^N$-grading on $\iGamma(M,E)$ and realises
it as an object in the category $\widetilde{\mathcal{V}}$. 
From the $\tilde\sigma$-equivariance of the bundle, the actions of
$\widetilde\TT^N$ on $C^\infty(M)$ and $\iGamma(M,E)$ are 
such that, for $r,r'\in \ZZ^N$, the products $f_r\psi_{r'}$ and
$\psi_{r'}f_r$ both belong to the $r+r'$-graded subspace of
$\iGamma(M,E)$, whence the result.\endproof

To simplify our formul{\ae}, we shall also use the notation $\E:=\iGamma(M,E)$ and $\A:=\Cinf(M)$. Then the vector space $\E$ is finitely generated and projective as a (right, say) $\A$-module. Now, $\TT^N$-equivariance of the $\A$-module $\E$ implies a crucial property of the corresponding projection that will prove invaluable throughout the paper ({\em cf}. \cite[Proposition 11.2.3]{blackadar}).

\begin{lem}
\label{le:p-cov}
There exists a finite-dimensional $\TT^N$-module $V$ such that the defining projection $\pp:V\otimes\A\to\E$ is a morphism in the category $\widetilde{\mathcal{V}}$.
\end{lem}

\proof 
The fact that $E$ is a torus-equivariant vector bundle means that there exists a finite dimensional $\TT^N$-module $\lambda:\TT^N\times V\to V$ such that $\E$ is $\TT^N$-equivariantly isomorphic to a direct summand of the $\A$-bimodule $V\otimes \A$ equipped with the diagonal $\TT^N$-action $\lambda\otimes\sigma$. Thus the projection $\pp:V\otimes \A\to\pp(V\otimes\A)\simeq \E$ is $\TT^N$-equivariant, as required.\endproof

As mentioned, the Hermitian structure on the bundle $E\to M$ is determined by a fibre metric, which amounts 
to a $C^\infty(M)$-valued Hermitian product on $\iGamma(M,E)$:
\begin{equation}\label{herm0}
\la\cdot,\cdot\ra:\E\times \E\to\A. 
\end{equation}
In light of Lemma~\ref{le:p-cov}, this is {\em expressed} as
\begin{equation}\label{herm1}
\la\phi,\psi\ra:=\sum_j \phi_j^*\psi_j,
\end{equation}
where we write 
$\phi=\sum_j e^j\otimes \phi_j$ and $\psi=\sum_j e^j\otimes \psi_j$ with respect to a choice of orthonormal basis $\{e^j\}_{j=1}^n$ for the $n$-dimensional space $V$ on which $\TT^N$ acts by unitaries. 
 
Next we use the Hermitian structure for a collection of new norms on the space $\E$ that, we stress, are different from the family of Fr\'{e}chet semi-norms on $\E$ mentioned before.

\begin{defn}\label{de:p-norm}
For each positive integer $p\geq 1$, the $p$-norm $\|\cdot\|_p$ on $\E$ is defined by
\begin{equation}\label{p-norm}
\|\cdot\|_p:\E\to\C,\qquad \|\phi\|_p:=\left( \int_M \la\phi,\phi\ra^{p/2}\right)^{1/p} 
\, , \qquad \textup{for}~~ \phi\in \E \, .
\end{equation}
The $C^*$-norm on $\E$ is defined to be
\begin{equation}\label{inf-norm}
\|\cdot\|:\E\to\C,\qquad \|\phi\|:=\|\la\phi,\phi\ra\|^{1/2} \, , \qquad \textup{for}~~ \phi\in \E \, ,
\end{equation}
where the norm $\|\cdot\|$ on the right hand side is the $C^*$-norm on $\A$ defined in Eq.~\eqref{C-norm}.
\end{defn}

Of course, in this definition we have used the fact that $\la\phi,\phi\ra$
is a positive element of the algebra $\A$ (by which we mean that it is positive when viewed in the $C^*$-completion) in order to define the
square root $\la\phi,\phi\ra^{1/2}$. Immediately we obtain the
following result.

\begin{lem}\label{le:norm-mor}
The maps $\|\cdot\|_p:\E\to\C$ and $\|\cdot\|:\E\to\C$ are morphisms in $\widetilde{\mathcal{V}}$.
\end{lem}

\proof The fact that the vector bundle $E$ is torus-equivariant
is equivalent to the statement that the inclusion of
$\A$-modules $\E\hookrightarrow V\otimes \A$ is
torus-equivariant and hence a morphism in the category
$\widetilde{\mathcal{V}}$. Since the product and $*$-structure on $\A$ are also morphisms in $\widetilde{\mathcal{V}}$, it is clear that the map
$\la\cdot,\cdot\ra:\E\to\A$ is a morphism in the category.
Since the integration map $f\mapsto \int_M f$ is also a morphism
in $\widetilde{\mathcal{V}}$, it follows that the norms $\|\cdot\|_p$ and $\|\cdot\|$ are constructed as
compositions of such morphisms, whence the result.\endproof

These arguments conveniently place the theory of vector bundles and normed vector spaces of sections in our categorical framework. There is  a parallel theory of torus-equivariant principal bundles over $M$, which we now describe in simple categorical terms. 

Let $G$ be a compact Lie group and let $P\to
M$ be a smooth principal $G$-bundle over $M$, so that $P$ carries a smooth right
$G$-action
\begin{equation}\label{prinact}
\Phi:P\times G\to P, \qquad (p,g)\mapsto \Phi_g(p),
\end{equation}
with quotient space $M\simeq P/G$. Let $c:\widetilde \TT^N\to \TT^N$
be a covering torus and assume the action $\sigma$ of $\TT^N$
on $M$ can be lifted to a smooth isometric action $\tilde \varsigma$ of $\widetilde\TT^N$ on $P$ 
commuting with the action of $G$. Then we say that $P$ is a {\em
$\tilde\varsigma$-equivariant principal $G$-bundle} over $M$.

\begin{lem}\label{le:prin} The vector space $C^\infty(P)$ is an algebra in the
category $\widetilde{\mathcal{V}}$. The canonical algebra inclusion
$C^\infty(M)\hookrightarrow C^\infty(P)$ is a morphism in the
category.\end{lem}

\proof The first claim follows in exactly the same way as
Lemma~\ref{le:grad}. The second claim
follows from the very definition of $P$ being
$\tilde\varsigma$-equivariant.\endproof

As mentioned, principal and vector bundles over $M$ are very much `parallel' theories, since it is possible to pass rather easily from one to the other, as we shall now recall. Let $\rho:G\to\End(V)$ be a finite-dimensional representation of $G$ on a complex vector space $V$. 
Then there is a vector bundle over $M$ associated to the representation $\rho$, defined by
$$
E=P\times_G V:=\left\{(p,v)\in P\times V~|~(\Phi_g(p),v)=(p,\rho(g^{-1})v)\right\}. 
$$ 
It is a classical result that there is an isomorphism
\begin{equation}\label{assocsec}
\iGamma(M,E)\simeq \Cinf(P)\boxtimes_\rho V,
\end{equation}
where we write 
\begin{equation}
\Cinf(P)\boxtimes_\rho V:=\left\{ \phi\in
C^\infty(P)\otimes
V~|~(\Phi_g\otimes\id)(\phi)=(\id\otimes\rho(g^{-1})(\phi)\right\}
\end{equation}
for the space of smooth $G$-equivariant maps from $P$
to $V$. 

Furthermore, if $P$ is a $\tilde
\varsigma$-equivariant principal bundle, then $E$ is equivariant as well.
Indeed, one writes $[p,v]$ for the $G$-equivalence class of the point $(p,v)\in P\times V$. 
Then, $\tilde\sigma$-equivariance of $E$ is provided by the action
$$\tilde\sigma:\widetilde\TT^N\times E\to E,\qquad
\tilde\sigma([p,v]):=[\tilde\varsigma(p),v].
$$
This is well-defined, since the action $\tilde\varsigma$ on $P$
commutes with the $G$-action, so the definition of $\tilde\sigma$
does not depend on the choice of representative of the $G$-equivalence class.

The inverse of this construction,  allowing us to pass from vector bundles back to principal bundles, is recalled in the next result.
As before, let $c:\widetilde\TT^N\to \TT^N$ be a covering of the torus $\mathbb{T}^N$ and let $G$ be a compact Lie group.

\begin{prop}\label{pr:assoc} Let $E$ be a $\tilde\sigma$-equivariant vector bundle over
$M$ with structure group $G$. Then there exists a
$\tilde\varsigma$-equivariant principal $G$-bundle $P\to M$, unique
up to isomorphism, such that $E$ is an associated vector
bundle.\end{prop}

\proof By definition, the fact that $E$ has structure group $G$
means that there is a principal $G$-bundle $P\to M$, unique up to isomorphism, and a finite-dimensional representation $\rho:G\to\textrm{End}(V)$, unique up to
unitary equivalence, such that $E\simeq P\times_G V$ as an associated vector
bundle \cite{kn:fdg}. Moreover, both $E$ and $P$ are completely determined by the transition functions of $P$ and the representation $\rho:G\to\textrm{End}(V)$ from which, out of $\tilde\sigma$, we obtain an action $\tilde\varsigma$ of  $\widetilde\TT^N$ on $P$. It is true by definition that the respective actions $\tilde\varsigma$ and $\tilde\sigma$ of $\widetilde\TT^N$ on $P$ and $E$ each cover the action $\sigma$ of $\TT^N$ on $M$. Since $E$ is $\tilde\sigma$-equivariant, the transition functions are $\sigma$-equivariant, whence $P$ is $\tilde\varsigma$-equivariant as well.\endproof

\subsection{Riemannian spin manifolds}
Now we add some extra structure to the discussion by taking $(M,g)$ to be an $m$-dimensional Riemannian manifold as before, but now assumed in addition to be {\em spin}. As such, this means that the manifold $M$ comes equipped with a spin principal bundle $\Sigma\to M$ with structure group $\Spin(m)$, the simply connected double cover of the orthogonal group $\SO(m)$. We write 
$$
\rho_{\frac{1}{2}} :\Spin(m)\to\End(V_{\frac{1}{2}} )\qquad\text{and}\qquad \rho_1:\Spin(m)\to\End(V_1)
$$
respectively for the spinor representation and the vector representation of the group $\Spin(m)$ on the $2^{[m/2]}$-dimensional space $V_{\frac{1}{2}} $ and the $m$-dimensional space $V_1$. The spinor bundle $\cS$ and the cotangent bundle $\Lambda^1(M)$ over $M$ are the vector bundles associated to $\Sigma$ via these two representations.
Consequently, we have isomorphisms
\begin{equation}\label{ass-spin}
\iGamma(M,\cS)\cong \Cinf(\Sigma)\boxtimes_{\rho_{\frac{1}{2}} }V_{\frac{1}{2}} ,\qquad \Omega^r (M)\cong \Cinf(\Sigma)\boxtimes_{\rho_1}\wedge^r V_1,
\end{equation}
where $\Omega^r(M):=\iGamma(M,\Lambda^r(M))$ denotes the $C^\infty(M)$-bimodule of $r$-differential forms on $M$, the smooth sections of the $r$th exterior power $\Lambda^r(M)$ of the cotangent bundle $\Lambda^1(M)$.  

Since the torus $\TT^N$ acts on $M$ by isometries, it lifts to an action of a covering torus $c:\widetilde\TT^N\to\TT^N$ upon the spinor bundle $\cS$ and upon each of the exterior bundles $\Lambda^r(M)$, which together translate into actions of $\widetilde \TT^N$ on the spaces of sections $\Gamma^\infty(M,\cS)$ and $\Omega^r(M)$, just as in Eq.~\eqref{eq:torus-bundle}. 
Proposition~\ref{pr:vecobj} immediately implies that $\iGamma(M,\mathcal{S})$ and $\Omega^r(M)$ are $\Cinf(M)$-bimodules in the category $\widetilde{\mathcal{V}}$. 

A key property of the spinor bundle is that its space of sections $\Gamma^\infty(M,\cS)$ is a module over the vector space of differential forms via the so-called Clifford multiplication:
\begin{equation}\label{cl-cliff}
\gamma: \Omega^r(M) \otimes_{C^\infty(M)} \Gamma^\infty(M,\cS) \to \Gamma^\infty(M,\cS).
\end{equation}
Indeed, writing $\gamma_{V}: \wedge^r V_1 \to \End(V_{\frac{1}{2}} )$ for the usual Clifford multiplication, we have that $\gamma = \id \otimes \gamma_{V}$ with respect to the identifications \eqref{ass-spin}.

\begin{lem}
\label{le:D-equi}
The Clifford multiplication $\gamma$ is a morphism in the category $\widetilde{\mathcal{V}}$.
\end{lem}
\proof
We already remarked that $\Gamma^\infty(M,\cS)$ and $\Omega^r(M)$ are objects in $\widetilde{\mathcal{V}}$. The map $\gamma$ is $\widetilde\TT^N$-equivariant since the torus acts by isometries lifted to the spinor bundle.
\endproof

Next, the Levi-Civita connection of the Riemannian metric $g$ lifts to the {\em spin connection} on the spinor bundle $\cS$, 
\begin{equation}\label{spin-co}
\nabla_\cS : \Gamma^\infty(M,\cS) \to \Omega^1(M) \otimes_{C^\infty(M)}    \Gamma^\infty(M,\cS) .
\end{equation}

\begin{lem}
The spin connection 
$\nabla_\cS$ 
is a morphism in $\widetilde{\mathcal{V}}$.
\end{lem}
\proof This follows by combining the compatibility between the Levi-Civita connection and the metric with the isometric action of the torus $\widetilde \TT^N$.
\endproof

The {\it Dirac operator} $D$ on the space of smooth sections of the spinor bundle is defined to be the composition of the spin connection $\nabla_\cS$ with the Clifford multiplication $\gamma$, namely
$$
D:= \gamma \circ \nabla_\cS:\Gamma^\infty(M,\cS)\to\Gamma^\infty(M,\cS) .
$$
Immediately we see that the Dirac operator has the following property.

\begin{lem}
\label{le:dirac}
The Dirac operator $D$ is a morphism in $\widetilde{\mathcal{V}}$.
\end{lem}

\proof
This follows from $D$ being the composition of continuous  $\widetilde\TT^N$-equivariant maps.
\endproof

On the Hilbert space $\h:=L^2(M,\mathcal{S})$ of square-integrable sections, the operator $D$ extends to an (unbounded) self-adjoint linear operator, which we continue to denote by $D$.
Moreover, the action of smooth functions on spinors by pointwise multiplication gives a
representation $\pi: C^\infty(M) \to\mathcal{B}(\h)$
as bounded operators on $\h$. 
The $\widetilde\TT^N$-action on spinors extends to a representation of $\widetilde\TT^N$ on $\h$ by unitary operators
$U(s)$, $s\in \widetilde \TT^N$, which leave the Dirac operator invariant, in the sense that
\begin{equation}\label{dirac}
U(s)\,D\,U(s)^{-1}=D  \qquad \text{for all}\quad s\in \widetilde\TT^N,
\end{equation}
and is such that
\begin{equation}\label{imp}
U(s)\,\pi(f)\,U(s)^{-1}=\pi(\sigma_{c(s)}(f)) \qquad \textup{for all}~~ f\in C^\infty(M),~s\in\widetilde\TT^N.
\end{equation}

The triple $(\A,\h,D)$ is called the {\em canonical spectral triple} on the spin manifold $(M,g)$. It is a spectral triple in the sense of A.~Connes \cite{ac:book}. 
With $\A=C^\infty(M)$, the Clifford multiplication \eqref{cl-cliff} yields an isomorphism between the $\A$-bimodule $\Omega^1(M)$ of one-forms over $M$ and the $\A$-bimodule of Connes' one-forms $\Omega^1_D(\A)$, the latter being defined as the vector space of bounded operators on $\h$ given by 
$$
\Omega^1_D(\A) := \left\{ \sum_j a_j[D,b_j] ~|~ a_j,b_j \in \A \right\} .
$$

\begin{rem}
\label{rem:equivariant-st}
\textup{
It is worth noting that the space $\h$ is `too big' to be an object in the category $\widetilde{\mathcal{V}}$. Indeed, the space of $L^2$-sections does not appear to have a decomposition property as in \eqref{hdp}.
Nevertheless, one can continue to make use of the categorical approach to spin geometry by working with the dense subspace of smooth sections $\Gamma^\infty(M,\cS)$.
}
\end{rem}

\subsection{Isospectral deformations of toric manifolds}\label{se:def-tor}
The results of the previous sections allow us to deform much of the geometry on the Riemannian spin manifold $M$.

\begin{defn} We write $C^\infty(M_\theta):=\L_\theta(C^\infty(M))$ for the image of the algebra $C^\infty(M)$ of smooth functions on $M$ under the deformation functor $\L_\theta$.\end{defn}

As the notation suggests, we think of $\Cinf(M_\theta)$ as the algebra
of smooth functions on an underlying virtual noncommutative space
$M_\theta$, the {\em toric noncommutative manifold} obtained from $M$ by the deformation functor. The algebra $\Cinf(M_\theta)$ can be completed in a suitable operator norm to obtain a $C^*$-algebra, denoted $C(M_\theta)$, which we think of as the algebra of continuous functions on the virtual space $M_\theta$. 

\begin{rem}\textup{
As a word of warning we remark that, due to the close interaction between the product in the algebra $\Cinf(M_\theta)$ and its $C^*$-norm, the latter cannot simply be equal to the classical $\sup$-norm on $\Cinf(M)$. Thus, in contrast with the case of the smooth function algebra $\Cinf(M_\theta)$, the $C^*$-completion $C(M_\theta)$ cannot be directly identified with its classical counterpart $C(M)$ as a vector space and equipped with a deformed product. We refer to \cite{mr:def} for a full analysis of this construction.}
\end{rem}

\subsubsection{Vector bundles on $M_\theta$}

As a consequence of Proposition~\ref{pr:vecobj} for equivariant vector bundles 
(with corresponding decomposition of sections as in Eq.~\eqref{hdp}), we may also apply
the deformation functor to any $\tilde\sigma$-equivariant vector
bundle $E$ over $M$.

\begin{defn}
We write $\iGamma(M_\theta,E)=\L_\theta(\iGamma(M,E))$ for the image of the $C^\infty(M)$-bimodule of sections 
$\iGamma(M,E)$ under the deformation functor $\L_\theta$. 
\end{defn}

The vector space $\iGamma(M_\theta,E)$ is considered to be the space of smooth sections of a noncommutative vector bundle 
over $M_\theta$.  
As already mentioned, from the properties of $\L_\theta$ it is automatic that $\iGamma(M_\theta,E)$ is a 
Fr\'{e}chet bimodule over the Fr\'{e}chet algebra $\Cinf(M_\theta)$. 

\begin{prop}\label{pr:tensiso} Let $E$, $F$ be $\tilde\sigma$-equivariant vector bundles over
$M$. Then $E\otimes F$ is a $\tilde\sigma$-equivariant vector bundle over $M$
and there are isomorphisms of $C^\infty(M_\theta)$-bimodules
\begin{align*}
\iGamma(M_\theta,E\otimes F) & \simeq\iGamma(M_\theta,E)\otimes_{C^\infty(M_\theta)}\iGamma(M_\theta,F) \\
& \simeq\iGamma(M_\theta,F)\otimes_{C^\infty(M_\theta)}\iGamma(M_\theta,E).
\end{align*}
\end{prop}

\proof Let us write $\tilde\sigma_E$ and $\tilde\sigma_F$ for the
actions of $\widetilde\TT^N$ on $E$ and $F$ respectively. Then
$E\otimes F$ becomes $\tilde\sigma$-equivariant when equipped with
the tensor product action $\tilde\sigma_E\otimes\tilde\sigma_F$.
Equivalently, 
the \emph{right} $C^\infty(M)$-module $\iGamma(M,E)$ and the \emph{left} $C^\infty(M)$-module $\iGamma(M,F)$
are objects in the category $\widetilde{\mathcal{V}}$
and hence so is the tensor product $C^\infty(M)$-bimodule
$\iGamma(M,E)\otimes_{C^\infty(M)}\iGamma(M,F)$. The isomorphism
$$
\iGamma(M,E\otimes
F)\simeq\iGamma(M,E)\otimes_{C^\infty(M)}\iGamma(M,F)
$$
is thus $\widetilde\TT^N$-equivariant and hence a morphism in the
category $\widetilde{\mathcal{V}}$. Applying the deformation functor
$\L_\theta$ yields the first isomorphism as stated. The second isomorphism is given
by the braiding defined by
Eq.~\eqref{th-braid} in the twisted category $\widetilde{\mathcal{V}}_\theta$.\endproof

 From the proof of Lemma~\ref{le:norm-mor}, we know that the canonical Hermitian structure on the undeformed bimodule $\iGamma(M,E)$ is a morphism in the category $\widetilde{\mathcal{V}}$. The image of this morphism under the deformation functor $\L_\theta$ is precisely the canonical Hermitian structure on the bimodule $\E:=\iGamma(M_\theta,E)$. It is still defined by the formula \eqref{herm1}, {\em viz.}
\begin{equation}
\label{herm1-theta}
\la\cdot,\cdot\ra:\E\times\E\to\Cinf(M_\theta),\qquad \la\phi,\psi\ra:=\sum_j \phi_j^*\psi_j,
\end{equation}
again upon writing $\phi=\sum_j e^j\otimes \phi_j$ and $\psi=\sum_j e^j\otimes \psi_j$ for elements in $\E$, but now using the deformed product in the algebra $\Cinf(M_\theta)$. Similarly, the norm $\|\cdot\|_p:\E\to\C$ defined in Eq.~\eqref{p-norm} defines a morphism in $\widetilde{\mathcal{V}}_\theta$.

\begin{lem}\label{le:nc-p-norm}
The normed vector spaces $(\iGamma(M,E),\|\cdot\|_p)$ and $(\iGamma(M_\theta,E),\|\cdot\|_p)$ are isometrically isomorphic.\end{lem}

\proof 
The fact that the vector spaces $\iGamma(M,E)$ and $\iGamma(M_\theta,E)$ are isomorphic follows from the very definition of the deformation functor. 
The norm $\|\cdot\|_p$ on $\iGamma(M,E)$ is a morphism in the category $\widetilde{\mathcal{V}}$ which, under the deformation functor, becomes a norm on $\iGamma(M_\theta,E)$ which is a morphism in the category $\widetilde{\mathcal{V}}_\theta$.\endproof

This approach to quantisation also lends itself nicely to
obtaining 
a differential calculus 
over $M_\theta$. As before, we write $\Omega(M) := \Gamma^\infty(M, \Lambda(M))$
for the (canonical) smooth differential forms on
$M$, with $\Lambda (M)$ the exterior algebra of its cotangent bundle. 

\begin{prop}\label{pr:d-calc}
The vector space $\Om(M)=\oplus_{r=0}^m \Om^r(M)$ is a differential graded algebra in the category $\widetilde{\mathcal{V}}$. Its image $\Om(M_\theta):=\L_\theta(\Om(M))$ under the quantisation functor is a differential graded algebra in the category $\widetilde{\mathcal{V}}_\theta$.
\end{prop}

\proof As a special case of Proposition~\ref{pr:vecobj}, each of the vector spaces $\Om^r(M)$, $r\geq 0$, is a 
Fr\'echet $\Cinf(M)$-bimodule in $\widetilde{\mathcal{V}}$. Since the action of $\TT^N$ on $M$ is by diffeomeorphisms, it
commutes with the exterior derivative
$\D:\Omega^r(M)\to\Omega^{r+1}(M)$, so the latter is a morphism in the category. In this way, 
$\Omega(M)$ is a differential graded algebra in the category
$\widetilde{\mathcal{V}}$ and we can apply the deformation functor $\L_\theta$.
We denote the resulting object by $\Omega(M_\theta)$, which is a Fr\'echet
$C^\infty(M_\theta)$-bimodule for left and right actions
obtained by deforming the classical actions using the
functor $\L_\theta$.
The latter acts as the identity on the differential $\D$, making $\Omega(M_\theta)$ into a differential graded
algebra with respect to the undeformed differential $\D$.
\endproof

With $\A=C^\infty(M_\theta)$, the Hermitian structure \eqref{herm1-theta} on an 
$\A$-bimodule $\E$ then extends to an $\Omega(M_\theta)$-valued sesquilinear map on the product
$\E\otimes_\A\Omega(M_\theta)\times\E\otimes_\A\Omega(M_\theta)$, by
\begin{equation}\label{ext-herm}
\la\phi\otimes\omega,\psi\otimes\zeta\ra=(-1)^{|\phi\|\omega|}\omega^*\la\phi,\psi\ra\zeta
\end{equation}
for each $\phi,\psi\in\E$, $\omega,\zeta\in\Omega(M_\theta)$.

\begin{defn}
\label{defn:deRham}
The de Rham cohomology groups ${\rm H}^r_\dR(M_\theta)$ of $M_\theta$ are defined to be
$$
{\rm H}_\dR^r(M_\theta) := \frac{\ker \D: \Omega^r(M_\theta) \to  \Omega^{r+1}(M_\theta) }{
{\rm im }~ \D: \Omega^{r-1}(M_\theta) \to  \Omega^{r}(M_\theta)}
$$
for each $r=0,1,2,\ldots$, where we define $\Omega^{-1}(M_\theta):=0$.
\end{defn}

Since the exterior derivative is undeformed, each cohomology group ${\rm H}_\dR^r(M_\theta)$ is canonically isomorphic to its classical counterpart ${\rm H}_\dR^r(M)$. In particular we have ${\rm H}^r_\dR (M_\theta) = 0$ if $r >m$ and, in fact, spaces of forms $\Omega^r(M_\theta)=0$ for $r>m$, with
$\Omega^m(M_\theta)$ being one-dimensional and spanned by a unique
volume form $\upsilon$.

\subsubsection{Principal bundles on $M_\theta$}

From Lemma~\ref{le:prin} we can also quantise
$\tilde\varsigma$-equivariant principal bundles over $M$. We write
$C^\infty(P_\theta)$ for the quantisation of $C^\infty(P)$, 
the image of the total space algebra $C^\infty(P)$ under the functor
$\L_\theta$. The action of the group $G$ on $C^\infty(P)$ induced by the action on $P$ is a morphism in $\widetilde{\mathcal{V}}$ and so yields a smooth action
\begin{equation}\label{twgp}
\Phi:C^\infty(P_\theta)\times G\to C^\infty(P_\theta)
\end{equation}
of $G$ on $C^\infty(P_\theta)$ by $*$-automorphisms, for which the
invariant subalgebra is $C^\infty(M_\theta)$, the quantisation of the base space algebra $\Cinf(M)$. Using standard Hopf algebra theory, the group action \eqref{twgp} dualises to a right coaction
\begin{equation}\label{Gco}
\delta_R:C^\infty(P_\theta)\to 
C^\infty(P_\theta)\htimes C^\infty(G),\qquad \delta_R(p)=p\bz\otimes p\bo,
\end{equation}
written in Sweedler notation.
The differential $\D:C^\infty(P_\theta)\to\Omega^1(P_\theta)$ being
undeformed, the algebra inclusion $j:C^\infty(M_\theta)\hookrightarrow C^\infty(P_\theta)$
extends to an inclusion of differential forms $j:\Omega^1(M_\theta)\hookrightarrow\Omega^1(P_\theta)$. 
We write
$$
\Omega^1_{\textup{hor}}(P_\theta):=C^\infty(P_\theta)\ j\left(\Omega^1(M_\theta)\right)
$$
for the sub-bimodule of $\Omega^1(P_\theta)$ of horizontal one-forms
(these are the analogue of the one-forms on $P_\theta$ that have been
pulled-back from the base $M_\theta$).
On the other hand, the coaction \eqref{Gco} gives rise to a canonical map which generates the vertical one-forms,
$$
\textup{ver}:\Omega^1(P_\theta)\to \Cinf(P_\theta)\htimes\Omega^1(G),\qquad \textup{ver}(p\, \D p'):=(p\otimes 1)(\id\otimes\D)(\delta_R(p)),
$$
defined for each $p,p'\in C^\infty(P_\theta)$, where $\Omega^1(G)$ denotes the space of one-forms on the group $G$.
 Using the maps $j$ and $\textup{ver}$, there is canonical sequence,
\begin{equation}\label{canseq}
0\to\Omega^1_{\textup{hor}}(P_\theta)\xrightarrow{j}
\Omega^1(P_\theta)\xrightarrow{\textup{ver}}\Cinf(P_\theta)\htimes\Omega^1(G)\to 0 ,
\end{equation}
which relates the horizontal one-forms to the vertical one-forms. Note that we could have found a version of this sequence  
in terms of the action of $G$, yet we prefer to pass to the equivalent coaction of $\Cinf(G)$ since it is in this setting that we are able to have an explicit formula for the map $\textup{ver}$. In order to have a quantum principal bundle, we need the sequence \eqref{canseq} to be exact \cite{brmaj}.

\begin{lem} The canonical sequence \eqref{canseq} is exact.\end{lem}

\proof For a classical principal $G$-bundle $P$ over $M$ there is a canonical exact sequence
$$
0\to\Omega^1_{\textup{hor}}(P)\xrightarrow{j}
\Omega^1(P)\xrightarrow{\textup{ver}} 
C^\infty(P)\htimes\Omega^1(G)\to 0,
$$
by its very definition.  
It is clear that the kernel of the map $\textup{ver}$, which is
identified with $\Omega^1_{\textup{hor}}(P)$, is a left
$\widetilde{\TT}^N$-module with action given by restricting the $\widetilde{\TT}^N$-action
on $\Omega^1(P)$. The deformation functor is just the identity on
the underlying vector spaces of these modules and so it preserves
exactness of the sequence.\endproof

In this way, it makes sense to speak of $\tilde\varsigma$-equivariant
quantum principal $G$-bundles with total space $P_\theta$ and base
space $M_\theta$, using the 
algebra inclusion $C^\infty(M_\theta)\hookrightarrow
C^\infty(P_\theta)$, thought of as a quantum principal bundle in the
sense of \cite{brmaj}. Just as was the case in \S\ref{se:toruseq} for classical bundles, we
now find that it is possible to pass between principal bundles and
vector bundles on toric noncommutative manifolds via the
associated bundle construction.

\begin{defn} \label{ncvb}
Let $C^\infty(M_\theta)\hookrightarrow C^\infty(P_\theta)$ be
a $\tilde\varsigma$-equivariant principal $G$-bundle over
$M_\theta$. The noncommutative vector bundle associated to the
principal bundle $P_\theta$ by the representation $\rho:G\to\End(V)$
is the $\Cinf(M_\theta)$-bimodule
$$
\Cinf(P_\theta)\boxtimes_\rho V:=\{ \phi\in C^\infty(P_\theta)\otimes
V~|~(\Phi_g\otimes\id)(\phi)=(\id\otimes\rho(g^{-1}))(\phi)\}.
$$
\end{defn}

It is known ({\em cf.} \cite{brmaj}) that  $\Cinf(P_\theta)\boxtimes_\rho V$ is finitely generated and projective as a (right, say) $C^\infty(M_\theta)$-module, thus qualifying it as a module of sections. Furthermore, 
$\Cinf(P_\theta)\boxtimes_\rho V$ is automatically a Fr\'{e}chet
$C^\infty(M_\theta)$-bimodule, since the same is true in the classical case.

\begin{prop}
\label{pr:def-assoc} Let $\iGamma(M_\theta,E)$ be a $\tilde\sigma$-equivariant vector bundle over $M_\theta$ with structure group $G$. Then there exists a
$\tilde\varsigma$-equivariant quantum principal $G$-bundle $P_\theta$ over
$M_\theta$, unique up to isomorphism, and a finite dimensional representation $\rho: G\to \End(V)$ such that
$\iGamma(M_\theta,E)\simeq \Cinf(P_\theta)\boxtimes_\rho V$ is an
associated vector bundle.\end{prop}

\proof Using the equivariance of the various bundle structures, this
follows immediately from Proposition~\ref{pr:assoc} and the
functorial properties of the deformation.\endproof

\subsubsection{Spin geometry of $M_\theta$}
Having dealt with the deformation of vector bundles and principal bundles, we now use the functorial quantisation procedure to deform the canonical spectral triple $(C^\infty(M),\mathcal{H},D)$ associated to the $m$-dimensional Riemannian manifold 
$(M,g)$. 
To find a spectral triple on the toric noncommutative manifold $M_\theta$, we leave the Dirac operator $D$ and the Hilbert space $\mathcal{H}$ as they are
and obtain a representation of the algebra $C^\infty(M_\theta)$ on $\mathcal{H}$ by deforming that of $C^\infty(M)$ on $\mathcal{H}$. 

Indeed, as described at the end of \S\ref{se:quant}, the quantisation functor $\L_\theta$ gives rise to a deformed left action of $\Cinf(M_\theta)$ upon the space $\iGamma(M_\theta,\mathcal{S})$, now denoted
\begin{equation}\label{th-rep}
\pi_\theta:C^\infty(M_\theta)\to\mathcal{B}(\iGamma(M_\theta,\mathcal{S})),\qquad
\pi_\theta(f)\psi:=
\sum_{r,s} \chi(r,s)
\pi(f_r)\psi_s,
\end{equation}
where $\psi\in \iGamma(M_\theta,\mathcal{S})$ and $f\in C^\infty(M_\theta)$ are decomposed as a sum of homogeneous components $\phi=\sum_s \psi_s$ and $f=\sum_r f_r$, respectively. 
Since the functor $\L_\theta$ is the identity on objects and
morphisms in $\widetilde{\mathcal{V}}$, we find by Lemma~\ref{le:norm-mor} that the $L^2$-norm $\|\cdot\|_2$ on $\iGamma(M_\theta,\mathcal{S})$ is unchanged, although it is now viewed as a morphism in the category $\widetilde{\mathcal{V}}_\theta$. 

We write $\h=L^2(M_\theta,\mathcal{S})$
for the Hilbert space completion of $\iGamma(M_\theta,\mathcal{S})$ in the
$L^2$-norm $\|\cdot\|_2$. Indeed, Lemma~\ref{le:nc-p-norm} shows that $L^2(M_\theta,\cS)$ and $L^2(M,\mathcal{S})$ are isometrically isomorphic, justifying our notation. The extension of the $C^\infty(M_\theta)$-module structure \eqref{th-rep}
is a representation of $C^\infty(M_\theta)$ on $\h$ by bounded operators:
$$
\pi_\theta:C^\infty(M_\theta)\to\mathcal{B}(\h) .
$$

The same argument goes for the Dirac operator. By Lemma~\ref{le:dirac}, the functor $\L_\theta$ applied  to the Dirac operator $D$ yields a linear map from $\Gamma^\infty(M_\theta,\cS)$ to itself. More precisely, one finds via Lemma~\ref{le:D-equi} that the Clifford multiplication \eqref{cl-cliff} is deformed into a map
\begin{equation}\label{nc-cliff}
\gamma_\theta: \Omega^1(M_\theta) \otimes_{\A} \Gamma^\infty(M_\theta,\cS) \to \Gamma^\infty(M_\theta,\cS).
\end{equation}
Similarly, the Levi-Civita spin connection \eqref{spin-co} is deformed into a map 
\begin{equation}\label{q-lc}
\nabla_\cS: \Gamma^\infty(M_\theta,\cS) \to  \Omega^1(M_\theta) \otimes_{\A} \Gamma^\infty(M_\theta,\cS),
\end{equation}
although for the time being we refrain from calling it a connection, having yet to define what this means in the deformed case.
The composition $\gamma_\theta \circ \nabla_\cS$ is easily seen to coincide with $\L_\theta(D)$. Upon identifying the Hilbert spaces $L^2(M_\theta,\cS)$ and $L^2(M,\cS)$, this operator agrees with the classical operator $D$ and so we continue to write $D$ instead of $L_\theta(D)$.

\begin{prop}\label{pr:spec-tr}
The datum $(C^\infty(M_\theta),\h,D)$ constitutes a spectral triple over the toric noncommutative manifold 
$M_\theta$.
\end{prop}

\proof
In terms of the unitaries $U$ of Eqs.~\eqref{dirac} and \eqref{imp},  
as in \cite{cl:id} one can write
$$
\pi_\theta(f)= \sum_r f_r \, U ( \half r \cdot \theta ) .
$$
The representation of $\Omega^1_D( C^\infty(M))$ on $\Gamma^\infty(M,\cS)$ ({\em cf}. Lemma \ref{le:D-equi}) is similarly deformed, giving an action of $\Omega^1_D(C^\infty(M_\theta))$ upon $\Gamma^\infty(M_\theta,\cS)$. Again, this action extends in a way which makes $\Omega^1_D( C^\infty(M_\theta))$ act upon $\h$ by bounded operators. In particular, this means that elements of the form $[D,\pi_\theta(f)]$ for any $f \in C^\infty(M_\theta)$ act by bounded operators. 
\endproof

This construction of the triple $(C^\infty(M_\theta),\h,D)$ precisely reproduces the one of \cite{cl:id}, but now using a categorical framework. It is clear that the datum $(C^\infty(M_\theta),\h,D)$ is an isospectral deformation of
the classical Riemmanian geometry of $M$, in the sense that the spectrum of the Dirac operator $D$ on $M_\theta$ coincides
with the spectrum of the classical Dirac operator on $M$. Consequently, the spectral triple is $m^+$-summable  and there is a noncommutative integral defined on $\Cinf(M_\theta)$ as a Dixmier trace,
namely
\begin{equation}\label{dixtr}
\intdix  f :=\textup{Tr}_\omega(\pi_\theta(f)|D|^{-m}),
\end{equation}
where $f\in \Cinf(M_\theta)$ and $\pi_\theta(f)$ denotes its image as an operator on $\h$.

Recall from Lemma~\ref{le:grad} that the classical
integral on $M$ is a morphism in the category $\widetilde{\mathcal{V}}$. The
noncommutative integral \eqref{dixtr} is precisely the image
under the deformation functor of the classical integral and is
therefore a morphism in $\widetilde{\mathcal{V}}_\theta$.
This means also that the norms $\|\cdot\|_p$ on
$\E=\iGamma(M_\theta,E)$ can be written as
\begin{equation}\label{nc-norm}
\|\cdot\|_p:\E\to\C,\qquad \|\phi\|_p=\left(\intdix
\la\phi,\phi\ra^{p/2}\right)^{1/p} , \qquad \textup{for}~~ \phi\in \E . 
\end{equation}
Furthermore, for each
$T\in\End_\A(\E)$ the $C^*$-inequality $\la T \phi, T \phi \ra \leq \| T \| \la \phi, \phi \ra$ implies that every such 
$T$ acts continuously with respect to each of the $p$-norms $\|\cdot\|_p$.

\subsubsection{Hodge structure on $M_\theta$}
\label{sect:hodge}
Our final addition to the geometric structure of the toric noncommutative manifold $M_\theta$ is the deformed analogue of a Hodge structure on the algebra of differential forms. To do so, we observe that the classical spin principal bundle $\Sigma$ over $M$ is deformed by the quantisation functor into a quantum principal bundle determined by the algebra inclusion $\Cinf(M_\theta)\hookrightarrow \Cinf(\Sigma_\theta)$. We therefore have isomorphisms
\begin{equation}\label{spin-sec}
\iGamma(M_\theta,\cS)\cong \Cinf(\Sigma_\theta)\boxtimes_{\rho_{\frac{1}{2}} } V_{\frac{1}{2}} ,\qquad \Omega^r(M_\theta)\cong \Cinf(\Sigma_\theta)\boxtimes_{\rho_{1}} \wedge^rV_{1},
\end{equation}
in analogy with the classical case.

\begin{prop}\label{pr:hodge}
In terms of the isomorphisms \eqref{spin-sec} for forms, there is a Hodge operator $\star_\theta:\Omega^r(M_\theta)\to\Omega^{m-r}(M_\theta)$ 
defined by 
$$
\star_\theta:=\id \otimes *  
$$
with ordinary Hodge operator $*: \wedge^r V_1 \to \wedge^{m-r} V_1$ on the $m$-dimensional vector space $V_1$. 
\end{prop}

\proof
It is clear that, in the classical case, the Hodge operator $\star$ is a morphism in the category $\widetilde{\mathcal{V}}$ and that it acts as $\id \otimes *$ on the vector space $C^\infty(\Sigma) \boxtimes_{\rho_1} \wedge^r V_1$. The result follows by applying the deformation functor $\L_\theta$ to this morphism, yielding the required morphism $\star_\theta$ acting upon $C^\infty(\Sigma_\theta) \boxtimes_{\rho_1} \wedge^r V_1$ as stated ({\em cf}. Proposition~\ref{pr:def-assoc}).
\endproof

\section{Gauge Theory on Toric Noncommutative
Manifolds}\label{se:ins}

This section is devoted to a description of gauge theory on toric
noncommutative manifolds $M_\theta$. We introduce the
(infinite-dimensional) group of gauge transformations of a
noncommutative vector bundle and the corresponding algebra of
infinitesimal gauge transformations. We then recall the notion of a
(compatible) connection on a noncommutative vector bundle and investigate the
behaviour of these objects under gauge transformations. 

\subsection{The gauge group of a vector bundle}\label{se:gaugegp}
Recall that the group of gauge transformations of a Hermitian vector bundle
$E$ over $M$ is defined to be the group of all vector bundle
automorphisms of $E$ which cover the identity on $M$ and preserve the
metric on the fibres of $E$. In this section we construct the
analogue of such objects for vector bundles over toric noncommutative manifolds. For
simplicity we consider vector bundles with structure group $G=\U(n)$, since this will be the case of interest later on in the paper.

Let $E$ be a vector bundle over $M_\theta$. For brevity, we continue to write $\A:=\CMt$ and $\E:=\iGamma(M_\theta,E)$. Suppose that $E$ has structure group $G$, so that there is a principal $G$-bundle $P_\theta$ over $M_\theta$ to which $E$ is associated via  
a finite dimensional $G$-representation $\rho:G\to\End(V)$. 
Definition~\ref{ncvb} identifies its module of sections as
$$
\E = \iGamma(M_\theta,E)\simeq\Cinf(P_\theta)\boxtimes_\rho V.
$$
The algebra of (continuous) right $\A$-module endomorphisms of $\E$ is
$$
\textup{End}(\E):=\left\{ T:\E\to \E~|~T(\phi a)=T(\phi)a~~\text{for
all}~\phi\in \E,~a\in \A\right\}.
$$
Since $G=\U(n)$ is compact, the vector space $V$ comes equipped with a canonical inner
product with respect to which the representation $\rho$ is unitary.

In what follows, a distinguished role will be played by the $\A$-module coming from the defining representation 
$V=\C^n$ of $\U(n)$ and having rank $n$ (the rank of a finite projective $\A$-module could be defined as its 0-th Chern number). This right $\A$-module will be denoted by $\E_0$ and the group of its unitary endomorphisms
defines the $\U(n)$ gauge group. 

\begin{defn} \label{def:gaugegp}
The {\em gauge group} is defined as the group
\begin{equation}\label{unis}
\mathcal{G}(\E_0):=\{U\in \textup{End}(\E_0)~|~ U^*U=UU^*=\id_{\E_0}\}
\end{equation}
of unitary endomorphisms of $\E_0$, where the adjoint operation $U\mapsto U^*$ on $\End(\E_0)$ is the one
induced by the Hermitian structure \eqref{herm1} and the canonical inner product on $\C^n$.
\end{defn}

Recall that the {\em dual module} of $\E$, defined as 
$$
\E':=\{\eta:\E\to \A~|~\eta(\phi a)=\eta(\phi)a, ~\phi\in\E,~ a\in \A\} , 
$$
is (anti-)isomorphic to $\E$ via the map $\eta\mapsto \la
\eta,\,\cdot\,\ra$.  Moreover, there is an isomorphism
$$
\E'\simeq \Cinf(P_\theta)\boxtimes_{\rho'} V' 
$$
 of $\A$-bimodules between the dual bundle $\E'$ and the 
vector bundle $\iGamma(P_\theta,V')$ associated to the dual representation $(\rho', V')$.

On the endomorphism algebra $\End(V)\simeq V\otimes V'$, the adjoint action of $G$ is just 
the tensor product representation $\textup{ad}:=\rho\otimes\rho'$. 
Then, with $\Cinf(P_\theta)\boxtimes_{\ad} \End(V)$ the vector bundle over $M_\theta$ associated to the  
representation $\textup{ad}$,
we have the following result.

\begin{prop}\label{pr:endid}
There is an isomorphism of right $\A$-modules
$$
\textup{End}(\E) \simeq \Cinf(P_\theta)\boxtimes_{\ad} \End(V) .
$$ 
\end{prop}

\proof As a consequence of Proposition~\ref{pr:tensiso}, there is an
isomorphism
$$
\Cinf(P_\theta)\boxtimes_{\ad} \End(V)= \Cinf(P_\theta)\boxtimes_{\ad}
(V\otimes V')\simeq \E\otimes_{\A}\E'
$$
of right $\A$-modules. Moreover, since $\E$ is finitely generated
and projective as a right $\A$-module, there is an isomorphism
$\End(\E)\simeq \E\otimes_{\A}\E'$, whence the result.
\endproof

In this way, the endomorphisms of the module $\E$ can be understood as sections
of the noncommutative vector bundle $\End(\E)=\iGamma(M_\theta,\End(E))$ associated to the adjoint
representation on $\End(V)$, exactly as in the classical case. 
From the general scheme spelled out in \S3.1, the space $\End(\E)=\iGamma(M_\theta,\End(E))$ is a Fr\'{e}chet algebra. As a consequence, the group $\cG(\E_0)$ of gauge transformations is a closed subspace of $\End(\E_0)$. 

At the infinitesimal level, the sections of the skew-Hermitian endomorphism bundle are
$$
\End^\textup{s}(\E):=\left\{ T\in\End(\E)~|~\la
T\phi,\psi\ra+\la\phi,T\psi\ra=0\right\},
$$
also a closed subspace of $\End(\E)$. More explicitly, for the $\A$-module $\E_0$ we can identify $\End^\textup{s}(\E_0)$ with the space $\Cinf_\R(P_\theta)\boxtimes_{\ad}\mathfrak{u}(n)$,
where $\Cinf_\R(P_\theta)$ is the space of self-adjoint elements in $\Cinf(P_\theta)$ and $\mathfrak{u}(n)$ is the subalgebra of skew-symmetric elements in $\End(V)$
\cite[Corollary 13]{lvs:nitcs}. 
This leads to

\begin{defn}\label{def:infgau} The Lie algebra of {\em infinitesimal gauge
transformations} is given by 
$$
\iGamma_\RR(\textup{ad}(P_\theta)):=\End^{\textup{s}}(\E_0)\simeq \Cinf_\RR(P_\theta)\boxtimes_{\textup{ad}}\mathfrak{u}(n),
$$
a (Fr\'echet) Lie algebra whose Lie bracket is defined by the commutator in
$\textup{End}^{\textup{s}}(\E_0)$.
\end{defn} 

We also write $\iGamma(\textup{ad}(P_\theta)):=\Cinf(P_\theta)\boxtimes_{\textup{ad}}\mathfrak{u}(n)$
for the Fr\'{e}chet Lie algebra whose Lie bracket is defined by the commutator in
$\textup{End}(\E_0)$. The $\Cinf(M_\theta)$-bimodule $\iGamma(\textup{ad}(P_\theta))$ will be understood as the the space of complexified sections of the adjoint bundle.

Our thinking of $\iGamma_\RR(\textup{ad}(P_\theta))$ as the collection of infinitesimal
gauge transformations is justified by the use of the exponential map, which is defined in the following way.

\begin{prop}\label{pr:exp}
There is a map 
$\Exp:\iGamma_\RR(\textup{ad}(P_\theta))\to \mathcal{G}(\E_0)$ with the property that, for each $X \in \iGamma_\RR(\textup{ad}(P_\theta))$, the set $\{U_t := \Exp tX ~|~t \in \RR\}$ defines a differentiable one-parameter family of elements in $\mathcal{G}(\E_0)$ such that $X =(\partial U_t/\partial t)|_{t=0}=:\dot U$. Conversely, if $U_t$ is such a family, then $X= \dot{U}$ is an element of $\Gamma^\infty_\RR(\ad P_\theta)$.
\end{prop}

\proof View an infinitesimal gauge transformation $X$ in the Lie algebra $C^\infty_\RR(P_\theta) \boxtimes_{\ad} \mathfrak{u}(n)$ as an element $X^* = -X$ in the space $C^\infty(P_\theta) \boxtimes_{\ad} \M_n(\C)$. The latter is a pre-$C^*$-algebra in $C(P_\theta)\boxtimes_{\ad} \M_n(\C)$ because $C^\infty(P_\theta)$ is so in $C(P_\theta)$. The usual exponential map $\Exp$ is a continuous $S^1$-valued function on the compact spectrum of $X$. Accordingly, the continuous functional calculus on the pre-$C^*$-algebra $C^\infty(P_\theta) \boxtimes_{\ad} \M_n(\C)$ gives an element $\Exp X \in C^\infty(P_\theta) \boxtimes_{\ad} \M_n(\C)$ whose spectrum is contained in $S^1 \subset \C$. In other words, $\Exp X$ is unitary and hence an element in $\mathcal{G}(\E_0)$. 
\endproof 

We shall also need spaces of differential forms taking values in the adjoint bundle. In order to handle such objects, we let
$\textup{End}(\E\otimes_{\A}\Omega(M_\theta))$ denote the algebra of
all right $\Omega(M_\theta)$-linear endomorphisms of
$\E\otimes_{\A}\Omega(M_\theta)$.

\begin{lem}\label{le:tech}There is an isomorphism of right $\A$-modules
$$
\E\otimes_\A\Omega(M_\theta)\simeq \Omega(M_\theta)\otimes_\A\E
$$
and hence an identification
$$
\textup{Hom}_\A(\E,\E\otimes_{\A}\Omega(M_\theta))\simeq
\Omega(M_\theta)\otimes_\A\textup{End}(\E).
$$
\end{lem}

\proof We have already seen the vector bundle $\Omega(M_\theta)$ as an
associated vector bundle, whence Proposition~\ref{pr:tensiso} yields
the first isomorphism. The second isomorphism follows from the
first one, combined with the fact that $\End(\E)\simeq
\E\otimes_\A \E'$ as right $\A$-modules.\endproof

Having shown that our notation is unambiguous, we make the following definitions.

\begin{defn}\label{def:higher}
Let $\E=C^\infty(P_\theta)\boxtimes_\rho V$ be a vector bundle over
$M_\theta$. We define
\begin{equation}\label{higher}
\Omega^r(P_\theta,V):=\E \otimes_\A\Omega^r(M_\theta), \qquad \textup{with} \quad r\geq 0 ,
\end{equation}
for the space of $r$-forms with values in the bundle. In particular,
when $V=\mathfrak{u}(n)$ and $\rho$ is the adjoint representation,
we write
$$
\Omega^r(\textup{ad}(P_\theta)):=\iGamma(\textup{ad}(P_\theta))\otimes_\A\Omega^r(M_\theta),
\qquad \textup{with} \quad r\geq 0,
$$
for the space of $r$-forms with values in the adjoint bundle.
\end{defn}

\subsection{Connections on vector bundles} Next we briefly
recall the theory of connections on vector bundles over a
toric noncommutative manifold $M_\theta$. 

\begin{defn} Let $\E=\iGamma(M_\theta,E)$ be a vector bundle over $M_\theta$. With $\A=C^\infty(M_\theta)$, a {\em right connection} on $\E$ is a linear map
\begin{equation}\label{conndef}
\n:\E\otimes_{\A} \Omega^r(M_\theta)\to
\E\otimes_{\A} \Omega^{r+1}(M_\theta)
\end{equation}
defined for all $r\geq 0$ and obeying the right graded Leibniz rule
\begin{equation}\label{leib}
\n(\omega \zeta)=(\n\omega)\zeta+(-1)^r\omega(\D \zeta)
\end{equation}
for all $\omega\in\E\otimes_{\A}\Omega^r(M_\theta)$ and $\zeta\in
\Omega(M_\theta)$.
\end{defn}

\begin{rem}\textup{A {\em left connection} on $\E$ is defined similarly, as a linear map
$$
\n:\Om^{r}(M_\theta)\otimes_\A\E\to\Om^{r+1}(M_\theta)\otimes_\A\E,
$$ 
now obeying a left Leibniz rule. In this section we present everything using right connections although, in the rest of the paper, we shall freely use both left and right connections.}
\end{rem}
 
From the Leibniz rule it follows that the composition
$$
\n^2:=\n\circ\n:\E\otimes_{\A}\Omega^r(M_\theta)\to
\E\otimes_{\A}\Omega^{r+2}(M_\theta)
$$
is $\Omega(M_\theta)$-linear, {\em i.e.}
$\n^2(\omega\zeta)=(\n^2\omega)\zeta$ for all
$\omega\in\E\otimes_{\A}\Omega^r(M_\theta)$ and $\zeta\in
\Omega(M_\theta)$.
The restriction $F:=\n^2:\E\to \E\otimes_{\A}\Omega^2(M_\theta)$ is called the {\em curvature} of
the connection $\n$. The curvature is $\A$-linear, {\em i.e.} an element in 
$\textup{Hom}_\A(\E,\E\otimes_\A\Omega^2(M_\theta))$. 

As we shall see momentarily, on a projective module $\E$ there is always a (right) connection.
Given any two connections $\n_1$, $\n_2$ on $\E$, their difference is $\A$-linear, {\em i.e.}, 
$$
(\n_1-\n_2)(\phi a)=((\n_1-\n_2)(\phi))a
$$
for all $\phi \in \E$, $a\in\A$, and so 
$\n_1-\n_2$ is an element of $\textup{Hom}_\A(\E,\E\otimes_\A\Omega^1(M_\theta))$.
Thus the space
of all connections on $\E$ is an affine space modeled on the latter space.

Then, let $\E$ be a projective module (of finite type) over $\A$. If $\E$ is identified
as a direct summand of the free module $\C^N\otimes \A$ by the
projection $\pp\in \M_N(\A)$, so that there is an isomorphism $j:\E \to \pp(\C^N\otimes \A)$, one has the
{\em Grassmann connection}
$$
\n_0:\E \to\E\otimes_\A\Omega^{1}(M_\theta)
$$
defined by the composition of maps:
\begin{multline}\label{gr-con}
\E \xrightarrow{j}\pp(\C^N\otimes \A)\xrightarrow{\id\otimes\D} \C^N \otimes\Omega^{1}(M_\theta) 
\xrightarrow{\pp\otimes\id}\pp(\C^N \otimes\Omega^{1}(M_\theta))\xrightarrow{j^{-1}\otimes\id}\E \otimes_{\A}
\Omega^{1}(M_\theta).
\end{multline}
This is simply denoted $\n_0:=\pp\circ (\id\otimes\D)$ and extended, by Leibniz rule, to a map
$\n_0:\E\otimes_\A\Omega^r(M_\theta)\to\E\otimes_\A\Omega^{r+1}(M_\theta)$.
Any connection $\n$ on $\E$ can then be written as
\begin{equation}\label{aff}
\n=\n_0+\omega,\qquad \textup{for some} \quad 
\omega \in \textup{Hom}_\A(\E,\E\otimes_\A\Omega^1(M_\theta)).
\end{equation}

\begin{lem}\label{le:gr-inv}
Let $E$ be a $\sigma$-equivariant vector bundle over $M$. Then the Grassmann connection $\n_0:\iGamma(M,E)\to\iGamma(M,E)\otimes_{\Cinf(M)}\Om^1(M)$ on $E$ is a morphism in the category $\widetilde{\mathcal{V}}_N$. Its image under the deformation functor $\L_\theta$ is the Grassmann connection on the noncommutative vector bundle $\E=\iGamma(M_\theta,E)$.
\end{lem}

\proof By Lemma~\ref{le:p-cov}, $\iGamma(M,E)$ is $\TT^N$-equivariantly isomorphic to a direct summand of a free torus-equivariant $\Cinf(M)$-module $V\otimes \Cinf(M)$. Let us write 
$j:\iGamma(M,E)\to \pp(V\otimes \Cinf(M))$ for this isomorphism. The Grassmann connection $\n_0$ on $\iGamma(M,E)$ is defined by  composition of maps just as in \eqref{gr-con} ({\em mutatis mutandis}).
With $\n_0$ being defined as a composition of torus-equivariant maps, we may apply the deformation functor to it.
The functor acts as the identity on objects and morphisms, so the image of $\n_0$ is nothing other than the Grassmann
connection on $\E=\iGamma(M_\theta,E)$.
\endproof

\begin{defn}\label{def:connecs} A right connection $\n$ on $\E$ is said to be {\em compatible} with the Hermitian structure $\la\cdot,\cdot\ra$ on $\E$ if it satisfies
\begin{equation}\label{comp}
\la\n\phi,\psi\ra+(-1)^{|\phi|}\la\phi,\n\psi\ra=\D\la\phi,\psi\ra , \qquad \textup{for all} \quad 
\phi,\psi\in\E\otimes_\A\Omega(M_\theta). 
\end{equation}
We write $\mathcal{C}(\E)$ for the collection  of all compatible connections on $\E$.\end{defn}

The Grassmann connection $\n_0$ is always compatible with the
canonical Hermitian structure defined in Eq.~\eqref{herm1}; the (affine) space
$\mathcal{C}(\E)$ is therefore always non-empty.  For a general
connection \eqref{aff}, the compatibility condition \eqref{comp}
reduces to the requirement that
$$
\la\omega\phi,\psi\ra+\la\phi,\omega\psi\ra=0 , \qquad \textup{for all} \quad \phi,\psi\in\E ;
$$
equivalently that $\omega$ satisfies
$\omega^*=-\omega$. Consequently, for the rank $n$ $\A$-module $\E_0$, the affine space $\mathcal{C}(\E_0)$ of compatible connections may be identified with the vector space $\Om^1(\textup{ad}(P_\theta))$ of one-forms with values in the skew-adjoint endomorphism bundle of $\E_0$.

Now we come to the issue of gauge theory of connections, 
{\em i.e.} of how connections behave if we apply gauge
transformations to noncommutative vector bundles.

\begin{defn}\label{def:gauge-act} The gauge group $\cG(\E_0)$ of
Definition~\ref{def:gaugegp} acts on the space $\cC(\E_0)$ by
\begin{equation}\label{gaugeact}
\cG(\E_0)\times \cC(\E_0)\to \cC(\E_0),\qquad (U,\n)\mapsto \n^U:=U\n U^*.
\end{equation}
\end{defn}

Given a gauge transformation $U\in\cG(\E_0)$, by differentiating the identity $UU^*=\id_{\cG}$
the difference of connections $\n^U-\n$ can be written as
$$
\n^U-\n=U(\n U^*)=-(\n U)U^*.
$$
Expressing some other connection $\widetilde\n$ as $\widetilde\n=\n+\omega$ for $\omega\in\Om^1(\textup{ad}(P_\theta))$, we get
$$
\widetilde\n^U=U(\n+\omega)U^*=\n+U(\n U^*)+U\omega U^*.
$$
Then, writing $\widetilde\n^U=\n+\omega^U$, the transformation
rule for the matrix-valued one-form $\omega$ is 
\begin{equation}\label{newalpha}
(U,\omega)\mapsto \omega^U:=-[\n, U]U^*+U\omega U^*,
\end{equation}
giving an explicit formula for the action of $\cG(\E_0)$ upon $\cC(\E_0)$ when we identify the latter affine space with the vector space
$\Om^1(\textup{ad}(P_\theta))$.

The remainder of this section is dedicated to showing that any
connection $\n\in\cC(\E)$ may be lifted to a connection on the
corresponding endomorphism bundle $\End(\E)$ and to consequences of this lifting. 

To do so, we shall need the
map
\begin{align}
[\n, \,\cdot \; ]&:\textup{End}(\E\otimes_{\A}\Omega(M_\theta))\to\textup{End}(\E\otimes_{\A}\Omega(M_\theta)), \nonumber \\
\label{comm}[\n,T]&=\n\circ T -(-1)^{|T|}T\circ \n,
\end{align}
where $|T|$ denotes the degree of $T$ for the
$\ZZ$-grading of $\Omega(M_\theta)$.  
Using Lemma~\ref{le:tech} one straightforwardly checks ({\em cf.} \cite{teh}) that the map $[\n,\cdot\,]$ is right $\Omega(M_\theta)$-linear,
$$
[\n,T](\omega \rho)= \left([\n,T](\omega)\right)(\rho) \qquad \text{for all} \quad\omega\in\textup{End}(\E\otimes_{\A}\Omega(M_\theta))
,~\rho\in\Omega(M_\theta),
$$ 
and hence well-defined. Moreover, it is a graded
derivation, in the sense that
\begin{equation}\label{gr-leib}
[\n,S\circ T]=[\n,S]\circ T+(-1)^{|S|}S\circ[\n,T]
\end{equation}
for all $S,T\in\textup{End}(\E\otimes_{\A}\Omega(M_\theta))$. We use all of this to define a connection on the 
endomorphism bundle over $M_\theta$.

\begin{prop}\label{endcon}
The formula \eqref{comm} defines a right connection
$$ 
[\n, \,\cdot \; ]:\textup{End}(\E)\otimes_\A\Omega^r(M_\theta)\to\textup{End}(\E)\otimes_\A
\Omega^{r+1}(M_\theta)
$$
on the right $\A$-module $\End(\E)$. This connection has well-defined
restrictions
\begin{equation}\label{endocon}
[\n, \,\cdot \; ]: \Omega^r(\textup{ad}(P_\theta))\to\Omega^{r+1}(\textup{ad}(P_\theta)) ,
\quad 
[F, \,\cdot \; ]: \Omega^r(\textup{ad}(P_\theta))\to\Omega^{r+2}(\textup{ad}(P_\theta)) ,
\end{equation} 
with $F= \nabla^2 $ being the curvature of $\nabla$.
\end{prop}

\proof 
For each $a\in \A$ we can take $S:=a\cdot\mathbbm{1}_{\A}$, {\em i.e.} we view $a\in\A$ as an endomorphism of $\E\otimes\Om(M_\theta)$ by left multiplication. Then Eq.~\eqref{gr-leib} becomes a graded Leibniz rule for $[\n,\cdot\,]$, viewed as a right connection on the right $\A$-module $\End(\E)$.
The well-definedness of the restriction is easily established ({\em cf.} \cite{teh}).
\endproof
\noindent
The connection in Proposition~\ref{endcon} will play a crucial role later in the paper.

\subsection{Weitzenb\"ock formula on $M_\theta$}
\label{sect:weitzenboeck}

Next we derive a Weitzenb\"ock formula for Dirac operators with coefficients on toric noncommutative manifolds. 
To do so, we shall need the Clifford multiplication \eqref{nc-cliff}, {\em i.e.} the $C^\infty(M_\theta)$-bimodule map
$$
\gamma_\theta: \Omega^p(M_\theta) \otimes_\A \Gamma^\infty(M_\theta,\cS) \to \Gamma^\infty(M_\theta,\cS),
$$
together with the canonical connection \eqref{q-lc} on the spinor bundle $\cS$,
$$
\nabla_\cS:\Gamma^\infty(M_\theta,\cS)\to\Om^1(M_\theta)\otimes_{\A}\Gamma^\infty(M_\theta,\cS),
$$  
whose curvature we denote by $F_\cS$.
Let $\E = \Gamma^\infty(M_\theta,E)$ be a 
vector bundle over $M_\theta$ equipped with a connection $\nabla_\E:\E\to\E\otimes_{\A} \Om^1(M_\theta)$, 
with curvature denoted $F_{\E}$.

\begin{defn}
The {\it Dirac operator with coefficients in $\E$} is the composition 
$$
D_\E: \E \otimes_\A \Gamma^\infty(M_\theta,\cS) \xrightarrow{\nabla_{\E \otimes \cS}} 
\E \otimes_\A\Omega^1(M_\theta) \otimes_\A \Gamma^\infty(M_\theta,\cS) \xrightarrow{\id \otimes \gamma_\theta} \E \otimes_\A \Gamma^\infty(M_\theta,\cS), 
$$
where $\nabla_{\E \otimes \cS} :=\nabla_\E \otimes \id + \id \otimes \nabla_\cS$ is the tensor product connection. 
\end{defn}

To obtain our result, we shall need a formula for the Hilbert adjoint $\n^*_{\E\otimes\cS}$ of the connection $\n_{\E\otimes\cS}$, taken with respect to the inner product on $\E \otimes_\A \Omega^1(M_\theta) \otimes_\A \Gamma^\infty(M_\theta,\cS)$ induced by a Hermitian structure as in \eqref{ext-herm}. Indeed, since both the Hermitian structure and the Hodge operator $\star_\theta$ are the images under the deformation functor of their classical counterparts, one finds just as in the classical case that
$$
\nabla^*_{\E \otimes \cS} = (\id \otimes \star_\theta\otimes \id) \circ \nabla_{\E \otimes \cS}\circ (\id \otimes  \star_\theta \otimes \id) 
$$ 
as an operator on $\E \otimes_\A \Omega^r(M_\theta) \otimes_\A \Gamma^\infty(M_\theta,\cS)$. 

In classical geometry, the Weitzenb\"ock formula relates the square of a Dirac operator with coefficients to the curvature of the bundle connection. The next theorem establishes such a formula for connections on vector bundles over toric noncommutative manifolds.

\begin{thm}
\label{thm:weitzenbock}
The operator $D_\E:\E \otimes_\A \Gamma^\infty(M_\theta,\cS)\to \E \otimes_\A \Gamma^\infty(M_\theta,\cS)$ has square
\begin{equation}\label{weitz}
(D_\E)^2 = \nabla^*_{\E \otimes \cS} \nabla_{\E \otimes \cS}+ \gamma_\theta(F_{\cS}) + \gamma_\theta(F_{\E}) .
\end{equation}
\end{thm}

\proof
Suppose first that $\nabla_\E$ is the Grassmann connection on $\E$. In this case, all of the maps appearing in Eq.~\eqref{weitz} coincide with their classical counterparts, whence the claim follows from the classical Weitzenb\"ock formula (see for example \cite[Thm. 3.4.2]{jost}). 

To establish the general case, first of all we remark that the compatibility between Clifford multiplication $\gamma_\theta$, the spin connection $\n_\cS$ and the tensor product connection $\n_{\Omega^1\otimes \cS}$ is expressed, as a map from $\Omega^1(M_\theta) \otimes_\A \Gamma^\infty(M_\theta, \cS)$ to $\Gamma^\infty(M_\theta, \cS)$, by
$$
\gamma_\theta\circ (\id \otimes \gamma_\theta) \circ \nabla_{\Omega^1 \otimes \cS} = \gamma_\theta\circ \nabla_\cS \circ \gamma_\theta.
$$
Consequently, we can write 
$$
D_\E^2 = (\id \otimes \gamma_\theta) \circ (\id \otimes \id \otimes \gamma_\theta) \circ  \left(\nabla_\E \otimes \id + \id \otimes  \nabla_{\Omega^1 \otimes \cS} \right) \circ\left( \nabla_{\E} \otimes \id + \id \otimes \nabla_{\cS} \right).
$$
Similarly we have
$$
\nabla_{\E \otimes \cS}^2=\left(\nabla_\E \otimes \id + \id \otimes  \nabla_{\Omega^1 \otimes \cS} \right)\circ\left( \nabla_{\E} \otimes \id + \id \otimes \nabla_{\cS}\right).
$$
The latter map $\nabla_{\E \otimes \cS}^2$ takes values in $\E \otimes_\A \Omega^1(M_\theta) \otimes_\A \Omega^1(M_\theta) \otimes_\A \Gamma^\infty(M_\theta,\cS)$, not yet involving the two-forms $\Omega^2(M_\theta)$. For this, 
we project $\Omega^1(M_\theta) \otimes_\A \Omega^1(M_\theta)$ onto the subspaces which are respectively braided-symmetric and braided-anti-symmetric, {\em i.e.} the $\pm 1$ eigenspaces for the braiding operator \eqref{th-braid} on the tensor product $\Omega^1(M_\theta) \otimes_\A \Omega^1(M_\theta)$. 
With $\L_\theta$ being an isomorphism of braided tensor categories, there are isomorphisms
\begin{align*}
\left( \Omega^1(M_\theta) \otimes_\A \Omega^1(M_\theta) \right)_{\Psi-{\textrm{sym}}} 
&\simeq C^\infty(\Sigma_\theta) \boxtimes_{\Spin(m)} (V_1 \otimes_S V_1); \\
\left( \Omega^1(M_\theta) \otimes_\A \Omega^1(M_\theta) \right)_{\Psi-{\textrm{asym}}} &\simeq C^\infty(\Sigma_\theta) \boxtimes_{\Spin(m)} (\wedge^2 V_1) \simeq \Omega^2(M_\theta).
\end{align*}

\noindent
Now recall that $\n_\E=\n_0+\omega$ for some connection one-form $\omega$. After projection onto two-forms, the
Levi-Civita connection $\nabla_{\Omega^1}: \Omega^1(M_\theta) \to\Omega^1(M_\theta) \otimes_\A \Omega^1(M_\theta)$ coincides with the exterior derivative $\D$. This means that, as a map from $\E \otimes_\A \Gamma^\infty(M_\theta,\cS)$ to $\E \otimes_\A \Omega^2(M_\theta) \otimes_\A \Gamma^\infty(M_\theta,\cS)$, we have
$$
\nabla_{\E \otimes \cS}^2 = (\nabla_0^2 + \nabla_0\, \omega + \omega\, \nabla_0 + \omega^2) \otimes \id + \id \otimes  \nabla_\cS^2 = F_{\E} \otimes \id + \id \otimes F_{\cS}.
$$
Composing the latter expression with Clifford multipication yields the second and third terms in the right-hand-side of the desired expression \eqref{weitz}.

Concerning the braided-symmetric part $\Omega^1(M_\theta) \otimes_\A \Omega^1(M_\theta)_{\Psi-{\textrm{sym}}}$, we find that Clifford multiplication coincides with taking the inner product $V_1 \otimes_S V_1 \to \C$ in the typical fibre $V_1$. In other words, the contribution to $D_\E^2$ is precisely $\nabla_{\E \otimes \cS}^* \nabla_{\E \otimes \cS}$.
\endproof

\section{Equivalence Classes of Connections}
\label{se:eq-classes}

We have studied so far the gauge theory of connections on toric noncommutative manifolds at the level of smooth sections of vector bundles. In this section we add some extra structure, making the space $\cC(\E_0)$ of compatible connections into a Banach space. This will allow us to apply a range of functional analytic techniques in order to determine the manifold structure of the space $\cC(\E_0)/\cG(\E_0)$ of gauge equivalence classes of connections.

\subsection{Sobolev theory} 

Let $M_\theta$ be a toric noncommutative manifold of dimension $m$ 
and let $\iGamma(M_\theta,E)$ be the Fr\'{e}chet $\Cinf(M_\theta)$-bimodule of `$\infty$-differentiable sections' of a torus-equivariant vector bundle over $M_\theta$. 
In fact, we can define the `$r$-differentiable sections' as the Fr\'echet $C^\infty(M_\theta)$-bimodule $\Gamma^r(M_\theta,E)$, the image under the deformation functor $\L_\theta$ of the $\Cinf(M)$-bimodule $\Gamma^r(M,E)$ of $r$-differentiable sections of a vector bundle $E$ over $M$. We then have $\Gamma^\infty(M_\theta,E) = \cap_r\, \Gamma^r(M_\theta,E)$.

With shorthand notation $\A:=\Cinf(M_\theta)$ and $\E:=\iGamma(M_\theta,E)$, in order to define a
collection of Sobolev norms on the vector space $\E$, we fix a compatible connection $\nom$ on $\E$. In addition, we take a compatible connection on $\Omega^1(M_\theta)$, now considered as the finitely generated projective $C^\infty(M_\theta)$-bimodule $\Gamma^\infty(M_\theta,\Lambda^1)$, where $\Lambda^1 := \Lambda^1 (M)$ denotes the cotangent bundle on $M$. Consider the $k$-fold tensor product
$\Lambda^{\otimes k} = \Lambda^1 \otimes \cdots \otimes \Lambda^1$ of the vector bundle $\Lambda^1$.
For any $k \geq 0$, we have tensor product connections $\nabla_{(k)}$
$$
\nabla_{(k)} : \Gamma^\infty(M_\theta,E \otimes \Lambda^{\otimes k}) \to  \Gamma^\infty(M_\theta,E \otimes \Lambda^{\otimes k}) \otimes_\A \Omega^1(M_\theta) \simeq \Gamma^\infty(M_\theta,E \otimes \Lambda^{\otimes (k+1)}),
$$
having used Proposition~\ref{pr:tensiso}.
By composition, we obtain maps
$$
\nabla^k := \nabla_{(k-1)} \circ \cdots \circ \nabla_{(0)} : \Gamma^\infty(M_\theta,E) \to \Gamma^\infty(M_\theta,E \otimes \Lambda^{\otimes k}).
$$
This should not be confused with simply taking $k$ powers of a connection; the latter would vanish identically for $k$ sufficiently large.

\begin{defn}
\label{def:sobolev}
For each pair of integers $p\geq 1$ and $k\geq0$, the {\em Sobolev
$(p,k)$-norm} $\|\cdot\|_{p,k}$ on $\E=\iGamma(M_\theta,E)$ is defined by
$$
\|\phi\|_{p,k} := \left( \|\phi\|_p^p+\|\nabla\phi\|_p^p+\ldots+\|\nabla^k\phi\|_p^p \right)^{\frac{1}{p}},\qquad \phi\in\E,
$$
where $\|\cdot\|_p$ is the $p$-norm on $\iGamma(M_\theta,E \otimes \Lambda^{\otimes l})$ defined in Eq.~\eqref{nc-norm} for $l \leq k$.
\end{defn}

For $p=2$ and any $k \geq0$ these Sobolev norms are Hilbert norms, using the obvious inner product giving the norm \eqref{nc-norm}. 
The standard properties of the norm $\|\cdot\|_{p,k}$, {\em i.e.}  
its being positive-definite and obeying the triangle
inequality, follow from those of $\|\cdot\|_p$. 
We just need to check that these norms on the vector
space $\E=\iGamma(M_\theta,E)$ are well-defined.

\begin{lem}\label{le:sob-ind}
For each $p\geq 1$ and $k\geq 0$ the topology on $\E$ defined by the Sobolev norm $\|\cdot\|_{p,k}$ is independent of the choice of
connections on $\iGamma(M_\theta,E)$ and $\iGamma(M_\theta,\Lambda^1)$.
\end{lem}

\proof 
In terms of the above prescription, any other choice of connections on $\iGamma(M_\theta,E)$ and $\iGamma(M_\theta,\Lambda^1)$ would give rise to connections
$$
\nabla_{(k)}' : \Gamma^\infty(M_\theta,E \otimes \Lambda^{\otimes k}) \to  \Gamma^\infty(M_\theta,E \otimes \Lambda^{\otimes k}) \otimes_\A \Omega^1(M_\theta) \simeq \Gamma^\infty(M_\theta,E \otimes \Lambda^{\otimes (k+1)}).
$$
Consequently, the Leibniz rule implies that the difference $\nabla_{(k)}' - \nabla_{(k)} =: \alpha_{(k)}$ is an element in $\mathrm{Hom}_\A ( \E \otimes_\A (\Omega^1)^{\otimes k},  \E \otimes_\A (\Omega^1)^{\otimes {(k+1)}})$. It is not difficult to deduce from this that 
\begin{equation}
\label{eq:nabla-k}
(\nabla')^k = \nabla^k + \gamma^{(k)}_1 \nabla^{k-1} + \cdots + \gamma^{(k)}_{k-1} \nabla + \gamma^{(k)}_k
\end{equation}
for some $\gamma^{(k)}_i \in \mathrm{Hom}_\A ( \E \otimes_\A (\Omega^1)^{\otimes {(k-i)}},  \E \otimes_\A (\Omega^1)^{\otimes k})$. 
We check by induction that the corresponding Sobolev norms $\|\cdot\|'_{p,k}$ and $\|\cdot\|_{p,k}$ are equivalent. For $k=1$ we find
that
\begin{align*}
\|\n' \phi\|_p=\|\nabla \phi+\alpha_{(0)} \phi\|_p\leq
\|\n\phi\|_p+\|\alpha_{(0)}\| \,\|\phi\|_p \leq (1+\|\alpha_{(0)}\|) \| \phi\|_{p,1}
\end{align*}
and hence that
$$
(\|\phi\|'_{p,1})^p=\|\phi\|_p^p+\|\n'\phi\|_p^p\leq 
\left(1+(1+\|\alpha_{(0)}\|)^p\right)\|\phi\|_{p,1}^p,
$$
having used the boundedness of $\alpha_{(0)}$ in the $C^*$-norm $\|\cdot\|$. Interchanging the roles of $\n$
and $\n'$ in this calculation establishes the reverse
inequality and so shows that $\n$ and $\n'$ define equivalent
Sobolev $(p,1)$ norms. 

Now, assume by induction that the norms $\| \cdot \|'_{p,l}$ and $\| \cdot \|_{p,l}$ are equivalent for all $l < k$. 
Using Eq.~\eqref{eq:nabla-k} we derive
\begin{align*}
\| (\n')^k \phi \|_p \leq \| \nabla^k \|_p + \| \gamma^{(k)}_1 \nabla^{k-1} \phi \|_p + \cdots \| \gamma^{(k)}_k \phi \|_p
\leq \left( 1+ \| \gamma^{(k)}_1 \| + \cdots + \| \gamma^{(k)}_k \| \right) \| \phi\|_{p,k}.
\end{align*}
By the very definition of $\| \cdot \|'_{p,k}$ and using the induction step, this implies that there exists a constant $c_k$ such that
$$
(\| \phi \|'_{p,k})^p = (\| \phi \|'_{p,k-1})^p + \| (\n')^k \phi \|_p^p \leq c_k \| \phi\|_{p,k}^p.
$$
Finally, interchanging $\n$ with $\n'$ gives the reverse inequality, and the result follows. 
\endproof

\begin{defn}\label{de:sb}
For each $p\geq 1$ and $k\geq 0$, the Sobolev space $\E_{p,k}=\Gamma_{p,k}(M_\theta,E)$ is defined to be the completion of the vector space $\E$ in the norm $\|\cdot\|_{p,k}$.
\end{defn}

Of course, there is {\em a priori} no
reason at all why the completions $\E_{p,k}$ should
bear any relation to their classical counterparts. Although the quantisation of the space of smooth sections of a vector bundle is defined by the deformation functor, connections on noncommutative vector bundles
over $M_\theta$ are not torus-equivariant in general and so cannot simply be quantised functorially. The
Sobolev completions for a noncommutative vector bundle might therefore be different from 
their classical counterparts. Nevertheless, we do find the following remarkable result.

\begin{prop}
\label{prop:banach-homeo}
For each $p\geq1$ and $k\geq0$ there are homeomorphisms of Banach spaces 
$$\Gamma_{p,k}(M,E)\to\Gamma_{p,k}(M_\theta,E),\qquad \Gamma^r(M,E)\to\Gamma^r(M_\theta,E).$$
\end{prop}

\proof By Lemma~\ref{le:gr-inv}, the Grassmann connections on $\iGamma(M,E)$ and $\iGamma(M_\theta,E)$ coincide as linear transformations. Moreover, by Lemma~\ref{le:nc-p-norm}, the $L^p$-norms on
$\iGamma(M,E)$ and $\iGamma(M_\theta,E)$ are just the same. It follows that the Sobolev $(p,k)$-norms
defined by the Grassmann connections on $\iGamma(M,E)$ and
$\iGamma(M_\theta,E)$ are equal. However, by
Lemma~\ref{le:sob-ind}, the Sobolev norms are independent of the
choice of connection, whence each of the Sobolev norms on the quantised
space $\iGamma(M_\theta,E)$ is equivalent to its classical
counterpart on the space $\iGamma(M,E)$. As a result, the Banach space completions in the
classical and quantum cases must be homeomorphic. 
The second homeomorphism above is direct from the definition of $\Gamma^r(M_\theta,E)$ at the beginning of this section.
\endproof

This means  we have at our disposal all of the usual Sobolev embedding theorems, but now for our $m$-dimensional toric noncommutative manifold $M_\theta$. We state them as follows.

\begin{thm}\label{th:techs} Let $\E_{p,k}=\Gamma_{p,k}(M_\theta,E)$ be the Sobolev spaces
as in Definition~\ref{de:sb}. Then:
\item 
\hspace{0.25cm} (i) 
if $k-(m/p)>r+1$, there is a continuous inclusion
$$\Gamma_{p,k}(M_\theta,E)\subset \Gamma^r(M_\theta,E).$$
In particular, if $\phi\in\Gamma_{p,k}(M_\theta,E)$ for some fixed $p\geq 1$ and all $k\geq 0$, 
then $\phi$ is smooth.
\item 
\hspace{0.25cm} (ii) 
if $k\geq\ell$ and $k-(m/p)\geq \ell-(m/q)$ then there is a continuous
inclusion
$$
\Gamma_{p,k}(M_\theta,E)\subset\Gamma_{q,\ell}(M_\theta,E).
$$
\item 
\hspace{0.25cm} (iii) 
for all $p\geq 1$, $k\geq0$, the inclusion map
$$
\Gamma_{p,k+1}(M_\theta,E)\to\Gamma_{p,k}(M_\theta,E)
$$ 
is compact.
\end{thm}

\proof
From classical Sobolev embedding theorems there are continuous
inclusion maps $\Gamma_{p,k}(M,E)\subset \Gamma^r(M,E)$ and
$\Gamma_{p,k}(M,E)\subset \Gamma_{q,\ell}(M,E)$ whenever the indices satisfy the stated inequalities. Due to the subtle observations made in proving Proposition~\ref{prop:banach-homeo}, we may now deduce that the same is true of the noncommutative Sobolev completions. In the same way, statement (iii) is deduced from the Rellich-Kondrachov
theorem for Sobolev spaces ({\em cf}. 
\cite{aub} for full details of the classical versions of these results). 
\endproof

\subsection{Analytic aspects of $\U(n)$ gauge theory on $M_\theta$}
We specialize the above discussion to $\U(n)$ gauge theories. 
Also, in what follows we only need to consider $L^2$-norms on our function spaces, rather than the more general $L^p$-norms. Accordingly, we compress the notation slightly, making the following definition.

\begin{defn}
Let $\E:=\iGamma(M_\theta,E) \simeq C^\infty(P_\theta) \boxtimes_{\U(n)} V$ be the space of smooth sections of a noncommutative vector bundle of rank $n$ over $M_\theta$, where $V$ is any representation of $\U(n)$,  and let $\Om^r(P_\theta,V)=\E\otimes_\A \Om^r(M_\theta) $  ({\em cf}. Definition~\ref{def:higher}). We write $\Om^r_k(P_\theta,V)$ for the completion of $\Om^r(P_\theta,V)$ with respect to the norm $\||\cdot\|_{2,k}$.
\end{defn}

In particular, we need the space
$\Om^1_3(\ad(P_\theta))=\Omega^1_3(P_\theta,\mathfrak{u}(n))$. 
In the remainder of the paper, we lighten the notation and drop the subscript on $\E_0$ to simply write $\E$ for the $\A$-module $\Cinf(P_\theta) \boxtimes_\rho \C^n$ of rank $n$. 
Denote by $\mathcal{C}:=\mathcal{C}(\E)$ the space of compatible
connections on the bundle $\E$, which is thus modeled on $\Omega^1(\ad P_\theta)$. 
We extend this to a Banach space.

\begin{defn} We write $\cC_3$ for the affine space of connections $\n$ on $\E$ of the
form $\n=\nom+\alpha$ for some
$\alpha\in\Om^1_3(\ad(P_\theta))$.\end{defn}

In this way, the space $\cC_3$ is identified with the space $\Om^1_3(\ad(P_\theta))$: the norm $\|\cdot\|_{2,3}$  gives it the
structure of a Banach space. Next we show how to equip the gauge
group $\cG:=\cG(\E)$ of Definition~\ref{def:gaugegp} with a Banach
structure as well. Since $\cG$ is a group and not a vector space, we cannot directly
apply the above argument to endow it with Sobolev norms. However, we use the fact that $\cG$ was obtained as the subset
\begin{equation}\label{g-emb}
\cG=\left\{U\in \End(\E)~|~U^*U=UU^*=\id_{\E}\right\} \subset
\End(\E).
\end{equation}
From Proposition~\ref{endcon} we know that the connection $\nom$
induces a connection on $\End(\E)$ by the formula
$[\nom,\,\cdot\,]$. As in Definition \ref{def:sobolev}, given a connection on $\Gamma(M_\theta, \Lambda^1)$ we can define the Sobolev $k$-norm on $\End(\E)$ by
\begin{equation}\label{p-norm-end}
\|T\|_{2,k} := \left( \|T\|_2^2+\| \n T\|_2^2+\ldots+\| \n^k T\|_2^2 \right)^{\frac{1}{2}}
\end{equation}
for each $T\in\End(\E)$, and whose completion is the Sobolev space $\Om^0_k(\End(\E))$. Note that the latter is a Hilbert space for any $k \geq 0$.

\begin{defn} We write $\cG_4$ for the closure of $\cG$ in
$\Om^0_4(\End(\E))$ (via the
embedding \eqref{g-emb}) with respect to the Sobolev norm
$\|\cdot\|_{2,4}$.\end{defn}

From these completions we obtain the following results, placing our earlier construction of gauge theory on noncommutative vector bundles in the context of Banach spaces.

\begin{lem}\label{le:estim}
For every $l \geq 3$ and every $k \geq l$, there exists a constant $d_k$ such that 
$$
\|ST\|_{2,l}\leq d_k\|S\|_{2,k}\|T\|_{2,l}
$$ 
for all $S \in\Om_k(\End(\E))$ and $T \in \Om_l(\End(\E))$.
\end{lem}

\proof At the level of smooth sections, there is a continuous product 
in the Fr\'echet algebra $\End(\E\otimes_\A\Om(M_\theta))$. We need to check that this product extends to a continuous product on the Sobolev completions in an appropriate way. Without loss of generality, we may assume that the Sobolev norms are defined using the canonical connections on the bundles $\End E$ and $\Lambda^1$, which are in particular torus-equivariant. Consequently, the Sobolev norms coincide with their classical counterparts ({\em cf}. Proposition~\ref{prop:banach-homeo}), allowing us to reduce the proof to the classical case. Indeed, let us assume that 
$S \in\Om(\End(\E))$ and $T \in \Om(\End(\E))$ are deformations of elements $S^{(0)}$ and $T^{(0)}$ in $\Omega(M,\End(E))$. 
With Proposition~\ref{pr:vecobj}, together with the corresponding decomposition for sections as in \eqref{hdp},
we can decompose $S$ and $T$ as a sum of homogeneous elements, and we get
$$
\| ST \|_{2,l} \leq \sum_{r,s} \| S^{(0)}_r T^{(0)}_s \|_{2,l} \leq \sum_{r,s} \| S^{(0)}_r \|_{2,l}  \|T^{(0)}_s \|_{2,l} 
$$
having used the classical result that $\Omega_{2,l}(M, \End E)$ is a Banach algebra, {\it i.e.} that the required inequality applies for $k=l$ ({\em cf}. \cite[Theorem 5.2.3]{Ada75}). Then, with $\nabla$ the torus invariant connection on $\End (E) \otimes \Lambda$, we derive
\begin{align*}
\| S^{(0)} \|_{2,l}^2 &= \int_M \langle S^{(0)},  S^{(0)} \rangle  + \cdots + \int_M \langle \nabla^l S^{(0)},  \nabla^l S^{(0)} \rangle  \\
& = \sum_r \int_M \langle S^{(0)}_r,  S^{(0)}_r \rangle  + \cdots + \sum_r \int_M \langle \nabla^l S^{(0)}_r,  \nabla^l S^{(0)}_r \rangle = \sum_r \| S^{(0)}_r \|^2_{2,l},
\end{align*}
the last equality holding by torus invariance of the integral on $M$. The observation that $\| S\|_{2,l} = \| S^{(0)}\|_{2,l}$ completes the proof.
\endproof

Note that the infinitesimal gauge algebra
$\iGamma(\ad(P_\theta))$ also has a Banach space completion
$\Om^0_4(\ad(P_\theta))$ in the norm $\|\cdot\|_{2,4}$ (in our earlier
notation, it is nothing other than the Banach space
$\Omega^0_4(P_\theta,\mathfrak{u}(n))$). In this sense, we have also $L^2(\ad (P_\theta)) \simeq\Omega_0^0(\ad (P_\theta))$.

In order to introduce analytic structures, we briefly recall some of the details of differentiable maps between Banach spaces.
Let $V,W$ be Banach spaces, let $\mathcal{U}$ be an open subset of $V$ and let $\Psi:\mathcal{U}\to W$ be a continuous map. For each point $x\in \mathcal{U}$, the derivative $\D\Psi_x:V\to W$ of $\Psi$ at $x$ in the direction $v\in V$ (if it exists) is defined to be the limit
\begin{equation}\label{ban-lim}
 \D\Psi_x:V\to W,\qquad \D\Psi_x(v):=\textup{lim}_{t\to 0}\, t^{-1}\left( \Psi(x+tv)-\Psi(x)\right),
\end{equation}
where $t\in (0,\infty)$ is always taken to be sufficiently small to arrange that $x+tv\in \mathcal{U}$. The function $\Psi$ is said to be {\em continuously differentiable} on $\mathcal{U}$ if $\D\Psi_x(v)$ exists for all $x\in \mathcal{U}$ and all $v\in V$ and if the associated map 
$$
\D\Psi:\mathcal{U}\times V\to W,\qquad \D\Psi(x,v):=\D\Psi_x(v),\quad x\in \mathcal{U},~v\in V,
$$ 
is continuous. Moreover, $\Psi$ is said to be of {\em class $C^r$} if it is $r$ times continuously differentiable. Finally, we say that $\Psi$ is {\em smooth} if it is of class $C^r$ for all $r\geq 0$.

We use this notion of differentiability to introduce a manifold structure on $\mathcal{G}_4$. Recall that a {\em Banach manifold} is a topological space equipped with an atlas of coordinate charts taking values in a Banach space such that the coordinate transition functions are smooth maps \cite{lang}. By a Lie group modeled on a Banach space, we mean a Banach manifold whose underlying set is a group such that the product and inversion operations are smooth maps.

\begin{prop}\label{pr:banach} The Sobolev closure $\cG_4$ is
a Lie group modeled on the Banach space $\Om^0_4(\textup{ad}(P_\theta))$. 
The action of $\cG$ on $\cC$ in \eqref{gaugeact} extends to a
continuously differentiable action 
\begin{equation}\label{sob-act}
\Psi:\cG_4\times\cC_3\to\cC_3, \qquad (U,\n)\mapsto U\n U^*,\qquad \textup{for} \quad U\in\cG_4,~\n\in\cC_3
\end{equation}
of $\cG_4$ on the space $\cC_3$ of compatible connections.\end{prop}

\proof 
We first establish the existence of an exponential map,
$$
\Exp:\Om^0_4(\textup{ad}(P_\theta))\to\mathcal{G}_4,
$$ 
as we had in the smooth setting ({\em cf.} Proposition~\ref{pr:exp}). Indeed, for any $X \in \Om^0_4(\textup{ad}(P_\theta))$ there exists a sequence $(X_n)$ with $X_n \in \Om^0(\textup{ad}(P_\theta))$ converging to $X$ in the $\|\cdot \|_{2,4}$-norm. Consequently, since
$$
(\lambda-X_n)^{-1} = (\lambda-X)^{-1} \left( 1+ (X-X_n)(\lambda-X)^{-1} \right)^{-1},
$$
the sequence of resolvents $((\lambda-X_n)^{-1})$ converges to $(\lambda-X)^{-1}$ in the same norm.
Then, $\Exp (X_n)$ can be defined as the Dunford integral 
$$
\Exp X_n = \frac{1}{2\pi i} \oint \Exp(\lambda ) (\lambda - X_n)^{-1} d \lambda,
$$
which is an element in $\cG$ for each $n$. It converges to $\Exp X$, defined through the same Dunford integral, which thus lies in the closure $\cG_4$ of $\cG$.

Just as it happens for finite-dimensional Lie groups, from the inverse function theorem this exponential map  is a homeomorphism in a neighbourhood of the origin and so it defines a local chart at the identity element $\id_{\mathcal{G}}$. Since $\mathcal{G}$ is dense in $\mathcal{G}_4$, left translation of this coordinate chart by elements of $\mathcal{G}$ provide a collection of coordinate charts covering all of $\mathcal{G}_4$. For $U_1,U_2\in\cG$, let $\mathcal{U}_1$, $\mathcal{U}_2$ be the coordinate charts centred at $U_1$, $U_2$ respectively. The transition function on the intersection of these patches is given by
left multiplication by the element $U_2U_1^*$. It is straightforward to verify that this map is smooth by repeatedly computing the limit \eqref{ban-lim} and then, using Lemma~\ref{le:estim}, to check continuity of the derivative. To establish the Lie group structure, it is sufficient to check that the map
$$
\cG_4\times\cG_4\to\cG_4,\qquad (U_1,U_2)\mapsto U_1U_2^*
$$
is smooth for all $U_1, U_2\in \cG_4$. This is also easily verified by computing the limit \eqref{ban-lim} and then using Lemma~\ref{le:estim} once again to check continuity of the derivative.

Finally, we turn to the group action
$\mathcal{G}\times\mathcal{C}\to\mathcal{C}$. Recall from Eq.~\eqref{newalpha} that the formula for the action of
the gauge group on the connection $\n:=\nom+\alpha$ is
\begin{equation}\label{g-alpha}
(U,\alpha)\mapsto \alpha^U:=-[\nom,U]U^*+U\alpha U^*.
\end{equation}
By Lemma~\ref{le:estim} there is a constant $d_3$ such that
\begin{align*}
\|\alpha^U\|_{2,3}&=\|U\alpha U^*-[\nom,U]U^*\|_{2,3}\leq \|U\alpha U^*\|_{2,3}+\|[\nom,U]U^*\|_{2,3}\\
&\leq d_3^2\|U\|^2_{2,3}\|\alpha\|_{2,3}+d_3\|U\|_{2,3}\|[\nom,U]\|_{2,3}\\
&\leq d_3^2\|U\|^2_{2,3}\|\alpha\|_{2,3}+d_3\|U\|_{2,3}\|U\|_{2,4},
\end{align*}
so that the action \eqref{g-alpha} is well-defined in the norm $\|\cdot\|_{2,3}$.  The space $\cC_3$ is in particular a Banach space and so it is a Banach manifold in the obvious way, with a single coordinate patch given by the identity map. In this way, the tangent space at any point in $\cC_3$ is identified with $\Om^1_3(\textup{ad}(P_\theta))$ itself. As mentioned above, the tangent space to a point $U\in\cG_4$ is identified with $\Om^0_4(\textup{ad}(P_\theta))$. One thus finds that the derivative as in \eqref{ban-lim} at the point $(U,\nom)\in\cG_4\times\cC_3$ and in the direction $(H,\alpha)\in\Om^0_4(\textup{ad}(P_\theta))\times \Om^1_3(\textup{ad}(P_\theta))$ is the map
\begin{equation}\label{fr-der}
\D\Psi_{(U,\nom)}(H,\alpha)=-[\nom,U_0H]+\alpha
\end{equation}
upon expressing $U=U_0\Exp(tH)$ in some local coordinate patch $\mathcal{U}_0$ centred at the point $U_0\in\cG$. Continuity of the map
$$
\D\Psi:\mathcal{U}_0\times\left(\Om^0_4(\textup{ad}(P_\theta))\times \Om^1_3(\textup{ad}(P_\theta))\right)\to\Om^1_3(\textup{ad}(P_\theta))
$$
is straightforward, whence the map $\Psi$ is continuously differentiable.\endproof

We finish the section with a string of technical results that we
shall call upon when needing them later on.
Classically they follows from the theory of elliptic operators.
Since in the noncommutative case no general theory
of ellipticity is available, we use in its place the more powerful methods
of perturbation theory for linear operators on Banach spaces.

\begin{lem}\label{le:bddrange}
\label{le:fred}
The linear operators
\begin{align*}
\n_\omega:\Om^r_k(\ad(P_\theta))&\to\Om^{r+1}_{k-1}(\ad(P_\theta))
\end{align*}
are Fredholm operators for all $r \geq 0$ and $k=3,4$.
\end{lem}
\proof
There exists $\omega\in \Om^1_3(\ad(P_\theta))$ such that
$\nom=\n_0+\omega$, where $\n_0$ is the Grassmann
connection on $\E$. By Lemma~\ref{le:gr-inv}, the latter 
is the same as its classical counterpart, which we
know defines a Fredholm operator as a map $\Om^r_k(\ad(P_\theta))\to\Om^{r+1}_{k-1}(\ad(P_\theta))$ for $k=3,4$.

Then, if $k=3$  the map
$\omega:\Om^r_3(\ad(P_\theta))\to\Om^{r+1}_3(\ad(P_\theta))$ is bounded by Lemma \ref{le:estim}. Thus, composing it with the compact map $\Om^{r+1}_3(\ad(P_\theta))\to\Om^{r+1}_2(\ad(P_\theta))$ ({\em cf}. Theorem~\eqref{th:techs}(iii)) again gives a compact map. It follows that $\nom: \Om^r_3(\ad(P_\theta))\to\Om^{r+1}_2(\ad(P_\theta))$ is a compact perturbation of the Fredholm operator $\n_0$ and is therefore itself Fredholm. 

When $k=4$ the desired result follows upon pre-composing the bounded map $\omega:\Om^r_3(\ad(P_\theta))\to\Om^{r+1}_3(\ad(P_\theta))$ with the compact embedding $\Om^r_4(\ad(P_\theta))\to\Om^r_3(\ad(P_\theta))$.
\endproof

The next result hinges on the following fact \cite[Theorem~VII.2.4]{kato}
from perturbation theory. Let $\h$ be a Hilbert space and
let $\Delta:\h\to\h$ be a self-adjoint linear operator with compact
resolvent and domain $\mathfrak{D}(\Delta)$. Suppose $T:\h\to\h$ is an operator 
such that 
\begin{equation}
\label{crit-A}
\tag{A}
\| T \phi\| \leq h ( \| \phi\|, \| \Delta \phi \|)  
 \qquad  \forall \phi \in \mathfrak{D}(\Delta) , 
\end{equation}
for some non-negative function $h(s,t)$ that is positive-homogeneous and monotonically increasing in both variables (Remark~VII.2.11 {\it loc. cit.}).
Then, the linear operator $\Delta+ x T$, for $x \in \RR$, is self-adjoint with domain $\mathfrak{D}(\Delta)$; it has compact resolvent if $|x| <  (h(0,1))^{-1}$.

\begin{lem}\label{le:comres}
The linear operator 
$$
\Delta_\omega:L^2(\ad(P_\theta))\to L^2(\ad(P_\theta)),
\qquad \Delta_\omega:=\nom^*\nom,
$$ 
is self-adjoint and has compact resolvent.
\end{lem}

\proof Take $\omega\in \Om^1_3(\ad(P_\theta))$ such
that $\nom=\n_0+\omega$, with $\n_0$ the Grassmann connection. Then
$$
\Delta_\omega=\Delta_0+\omega^*\n_0+(\n_0^*+\omega^*)\omega,
$$
where $\Delta_0=\n_0^*\n_0$ coincides with the classical
Laplacian associated to the Grassmann connection, which
therefore has the same spectrum and hence it has compact
resolvent. Furthermore, on the domain $\mathfrak{D}(\Delta_0)$ of $\Delta_0$ one can write
$$
(\n_0^*+\omega^*)\omega = \star_\theta (\n_0+\omega) \star_\theta \omega = \star_\theta [\nabla_0, \star_\theta \omega] + \omega \nabla_0 + \star_\theta\omega \star_\theta\omega,
$$
with $\star_\theta$ the Hodge star operator. Note that the latter is isometric with respect to the inner product on $\Omega^p(M_\theta)$, so that 
\begin{align*}
\| \omega^* \nabla_0 \phi \|_2 &\leq  \| \omega^* \| \| \nabla_0 \phi \|_2 \leq   \| \omega^* \| \| \Delta_0 \phi \|_2^{1/2} \| \phi \|_2^{1/2}
\intertext{and}
\| (\n_0^*+\omega^*)\omega \phi \|_2 &\leq \| [\nabla_0, \star_\theta \omega] \| \| \phi\|_2 + \| \omega\| \| \nabla_0 \phi\|_2 + \| \omega^*\omega \| \| \phi\|_2\\
& \leq \| [\nabla_0, \star_\theta \omega] \| \| \phi\|_2 + \| \omega\| \| \Delta_0 \phi \|_2^{1/2} \| \phi \|_2^{1/2}+ \| \omega^* \omega \| \| \phi\|_2.
\end{align*}
We conclude that $\Delta_\omega$ is a perturbation of $\Delta_0$ for which criterion \eqref{crit-A} holds with 
$$
h(s,t) = 2  \| \omega^* \|  (st)^{1/2} + \left( \| [\nabla_0, \star_\theta \omega] \| +  \| \omega^* \omega \| \right) s
$$
Since $h(0,1) = 0$, $\Delta_\omega$ is self-adjoint with domain $\mathfrak{D}(\Delta_0)$  and has compact resolvent. 
\endproof

\begin{prop}\label{pr:kernabla}
The kernel $\textup{Ker}\,\nom$ 
consists of
smooth sections
and there exists a constant
$c_k$ such that
\begin{equation}\label{bdd}
\|\phi\|_{2,k}\leq c_k\|\nom\phi\|_{2,k-1}
\end{equation}
for all $\phi\in\Om^0_k(\ad(P_\theta))$ such that
$\phi\,\bot\,\textup{Ker}\,\nom$.
\end{prop}

\proof If $\nom \phi=0$ then $\nom \cdots \nom\phi=0$ for an arbitrary
number of applications of $\nom$. Thus $\phi\in\Omega^0_k(\ad(P_\theta))$
for
all $k\geq0$ and $\phi$ is smooth by Theorem~\ref{th:techs}(i). 
Also, from the equality
$\|\phi\|_{2,k}^2=\|\phi\|_{2}^2+\|\nom\phi\|_{2,k-1}^2$, the claim \eqref{bdd} is equivalent to 
requiring that there is a constant $c>0$ for which
$$
\|\phi\|_{2}\leq c\|\nom\phi\|_{2}
$$
for all $\phi\in\Om^0_k(\ad(P_\theta))$ such that $\phi\,\bot\,\textup{Ker}\,\nom$. A suitable choice is
$c=\lambda^{-1}$, where
$$
\lambda=\textup{inf}
\left\{ (\|\phi\|_2)^{-1} \|\nom\phi\|_2 ~|~ \phi\,\bot\,\textup{Ker}\,\nom \right\},
$$
provided we show that $\lambda>0$. Since $(\Delta_\omega\phi,\phi)_2=(\nom\phi,\nom\phi)_2=\|\nom\phi\|^2_2$, 
such $\lambda$ is greater than the square root of the smallest
non-zero eigenvalue of $\Delta_\omega$ on $\Omega_0^0(\ad (P_\theta)) \simeq L^2(\ad (P_\theta))$. By
Lemma~\ref{le:comres}, the operator $\Delta_\omega$ has compact
resolvent, meaning in particular that zero is not an accumulation point of
its spectrum and hence that $\lambda>0$.\endproof

\subsection{Manifold structure of the quotient space} 
Having introduced the Sobolev completions $\cC_3$ and $\cG_4$ of the space of compatible connections and of the gauge group, we move to describe the quotient space $\cC_3/\cG_4$ of compatible connections modulo gauge equivalence.

Let $\n$ be a compatible connection on the bundle $\E$. The {\em isotropy group} $\Gamma^\n$ of $\n$ is 
the subgroup  of the gauge group $\mathcal{G}_4$ defined by
$$
\Gamma^\n:=\left\{  U\in \mathcal{G}_4~|~ U\n U^*=\n\right\} .
$$
As the following lemma shows, the isotropy group of any given connection is always non-trivial: there is a certain subgroup of $\mathcal{G}_4$ which fixes every point $\n$ of the space $\mathcal{C}_3$.

\begin{lem}\label{le:iso}
There exists a subgroup $\Gamma_0$ of $\mathcal{G}_4$ such that $\Gamma_0\subseteq\Gamma^\n$ for all compatible connections $\n$ in $\mathcal{C}_3$.
\end{lem}

\proof This hinges on the observation that, at the infinitesimal level, we have the direct sum decomposition of $\Om^0_4(\textup{ad}(P_\theta))$ into
\begin{align*}
\Om^0_4(\textup{ad}(P_\theta))& =\left(\Om^0_4(P_\theta)\boxtimes_{\mathbbm{1}}\mathfrak{u}(1)\right)\oplus \left(\Om^0_4(P_\theta)\boxtimes_{\textup{ad}}\mathfrak{su}(n)\right). 
\end{align*}
Let us focus on the summand $\Om^0_4(P_\theta)\boxtimes_{\mathbbm{1}}\mathfrak{u}(1)$, noting that if $H\in\Om^0_4(P_\theta)\boxtimes_{\mathbbm{1}}\mathfrak{u}(1)$ is such that $[\n,H]=0$, then $[\n^k,H]=0$ for all $k\geq 1$ and so $H\in\Om^0_k(P_\theta)\boxtimes_{\mathbbm{1}}\mathfrak{u}(1)$ for all $k\geq 0$
, whence $H$ is smooth by Theorem~\ref{th:techs}(i). The identification
$$
\iGamma(P_\theta)\boxtimes_{\mathbbm{1}}\C\simeq \Cinf(M_\theta)
$$
means that $H$ is in fact a self-adjoint element of $\Cinf(M_\theta)$ satisfying 
$[\n,H]=\D H=0$, {\em i.e.} it is a constant function on $M_\theta$ with values in the Lie algebra $\mathfrak{u}(1)$. Exponentiating this Lie algebra gives a subgroup $\Gamma_0$ of $\mathcal{G}_4$ which is necessarily a subgroup of $\Gamma^\n$.\endproof

The non-triviality of the isotropy group of each point of $\mathcal{C}_3$  means that, in order to obtain a nicely-behaved space of orbits of $\mathcal{G}_4$ in $\mathcal{C}_3$, we must quotient the subgroup $\Gamma_0$.

\begin{defn}
We write $\widetilde{\mathcal{G}}_4:=\mathcal{G}_4/\Gamma_0$ for the quotient of the gauge group $\mathcal{G}_4$ by the isotropy subgroup $\Gamma_0$ identified in Lemma~\ref{le:iso}.
\end{defn}

 Of course, the isotropy group $\Gamma^\n$ of a given connection $\n$ may be larger than the subgroup $\Gamma_0$, so we also need to discard the connections for which this happens and focus on the subspace of $\mathcal{C}_3$ on which the action of the quotient gauge group $\widetilde{\mathcal{G}}_4$ is indeed free.

\begin{defn}
A connection $\n$ in $\mathcal{C}_3$ is said to be {\em irreducible} if the isotropy group $\Gamma^\n$ is equal to $\Gamma_0 = \U(C^\infty(M_\theta))$, the unitary group of the algebra $C^\infty(M_\theta)$, otherwise it is said to be {\em reducible}. We write $\widetilde{\mathcal{C}}_3$ for the subset of $\mathcal{C}_3$ consisting of irreducible connections. 
\end{defn}

As shown in Proposition~\ref{pr:banach}, the tangent space to $\cG_4$ at the identity element $\id_{\cG}$ may be identified with the vector space $\Om^0_4(\ad(P_\theta))$. Similarly, the tangent space at each point of the quotient gauge group $\widetilde{\mathcal{G}}_4$ is identified with the quotient vector space
$$
\widetilde{\Om}^0_4(\ad(P_\theta)):=\Om^0_4(\ad(P_\theta))/\gamma_0,
$$
where $\gamma_0 = \mathfrak{u} (C^\infty(M_\theta))$ is the Lie algebra of the group $\Gamma_0$ identified in Lemma~\ref{le:iso}; it can be identified with the purely imaginary elements in the algebra ${\rm H}^0_\dR(M_\theta)$ of constant functions. Moreover, $\cC_3$ being an affine vector space, the tangent space at any of its points may be identified with the corresponding vector space $\Om^1_3(\textup{ad}(P_\theta))$. Since $\widetilde{\cC}_3$ is obtained from $\cC_3$ simply by deleting the reducible points, the typical tangent space of $\widetilde{\cC}_3$ is also equal to $\Om^1_3(\textup{ad}(P_\theta))$. These realisations of the tangent spaces of $\widetilde{\cC}_3$ and $\widetilde{\cG}_4$ will prove useful in obtaining the following result.

\begin{thm}\label{pr:charts}
The space $\cB=\widetilde{\cC}_3/\widetilde{\cG}_4$ is a Banach manifold with local charts
given by $\pi:\CO_{\omega,\ep}\to \cB$, where, for $\ep$ sufficiently small,
$$
\CO_{\omega,\ep}:=\left\{
\nom+\alpha~|~\alpha\in\Om^1_3(\ad(P_\theta))~\text{with}~\alpha\in\textup{Ker}\,\nom^* 
~\text{and}~\|\alpha\|_{2,3}<\ep\right\} .
$$
\end{thm}

\proof By Lemma~\ref{le:bddrange} we know that $\im \nom$ is a closed subspace of $\Omega^1_3(\ad(P_\theta))$, so there is an orthogonal
decomposition
\begin{equation}\label{splitting}
\Omega^1_3(\ad(P_\theta))=\im \nom\oplus\textup{Ker}\, \nom^*
\end{equation}
(in general this is only true if $\im \nom$ is replaced by its closure $\overline{\im \n}_\omega$). 
The subspace $\im \nom$ is identified with the tangent space to the orbit $\widetilde{\cG}_4(\nom)$ in $\widetilde{\cC}_3$ of the point $\nom$ under the action of the gauge group $\widetilde{\cG}_4$. 
Our first step is to construct a `local slice' for the action of $\widetilde{\cG}_4$. This is done as follows.

For $\CO_{\omega,\ep}$ as defined above, we consider the map
$$
\Psi:\widetilde{\cG}_4\times\CO_{\omega,\ep}\to\widetilde{\cC}_3,\qquad \Psi(U,\n):=U\n U^* , 
$$
that we show is a diffeomorphism in a
neighbourhood of $(\id_\mathcal{G},\nom)$. By the inverse function theorem, this would follow  were we 
to show that the derivative $\D\Psi$ obtained in Proposition~\ref{pr:banach} is an isomorphism at the point
$(\id_\mathcal{G},\nom)$. To this end,  
consider the infinitesimal neighbourhood of
$(\id_\mathcal{G},\nom)$ given by the spaces
$\widetilde{\Om}^0_4(\ad(P_\theta))$ and $\Om^1_3(\ad(P_\theta))$, so that
the
derivative of $\Psi$ is a map
$$
\D \Psi:\widetilde{\Om}^0_4(\ad(P_\theta))\times \textup{Ker}\,\nom^*\to
\Om^1_3(\ad(P_\theta)),
$$
with $\textup{Ker}\,\nom^* \subset \Om^1_3(\ad(P_\theta))$.
Using the formula \eqref{fr-der} for the
infinitesimal action of $U$ on a given connection
$\n=\nom+\alpha$, we have
$$
\D\Psi_{(\id_\mathcal{G},\nom)}(H,\alpha)=-[\nom,H]+\alpha.
$$
With respect to the splitting \eqref{splitting} of
$\Om^1_3(\ad(P_\theta))$, it is clear that the map $\D\Psi_{(\id_\mathcal{G},\nom)}$ is
surjective. Thus it is an isomorphism if and only if it is injective
which, by the very definition, 
is the case precisely when $\nom$ is irreducible.
Thus it follows from the inverse function theorem that there is a
neighbourhood $V_\omega$ of $\nom\in\widetilde{\cC}_3$ such that, with
$$
\widetilde{\cG}_{\ep}:=\{U\in \widetilde{\cG}_4~|~\|U-\id_{\widetilde{\cG}}\|_{2.4}<\ep\},
$$
the map $\Psi$ induces an isomorphism from $\widetilde{\cG}_{\ep}\times \CO_{\omega,\ep}\to
V_\omega$. In particular, this means that whenever $\n_1,
\n_2\in\CO_{\omega,\ep}$ are such that $U\n_1U^*=\n_2$ for some
$U\in\widetilde{\cG}_{\ep}$, we have $\n_1=\n_2$.

Finally we claim that this last conclusion is valid for any $U\in\widetilde{\cG}_4$ and not just for
those $U\in\widetilde{\cG}_{\ep}$. Let $\n_i=\nom+\alpha_i$, $i=1,2$, be a
pair of elements in $\CO_{\omega,\delta}$ for some $\delta>0$ (to be chosen), so that
$\|\alpha_i\|_{2,3}<\delta$. Then, the condition $U\n_1 U^*=\n_2$ is
equivalent to the condition $[\nom,U]=U\alpha_1-\alpha_2 U$. By Proposition~\ref{pr:kernabla}, for each $k\geq 0$ there exists a constant $c_k$ such that
$\|U-\id_{{\cG}}\|_{2,k}\leq c_k \|[\nom,U]\|_{2,k-1}$. Using this fact, we find that
\begin{align*}
\|U-\id_{\cG}\|_{2,1}&\leq c_1\|[\nom,U]\|_{2,0}=c_1\|U\alpha_1-\alpha_2
U\|_{2,0}\leq\left( \|U\alpha_1\|_{2,0}+\|\alpha_2 U\|_{2,0}\right) \\
&=c_1\left( \|\alpha_1\|_{2,0}+\|\alpha_2\|_{2,0}\right) \leq 2c_1\delta.
\end{align*}
Now with the estimates $\|U\alpha_i\|_{2,k}\leq d_k\|U\|_{2,k}\|\alpha_i\|_{2,3}$, using 
a `bootstrapping' argument one finds that, for each $k=0,\ldots,3$, there is a constant $c_k'$ such that 
$$
\|U-\id_{\cG}\|_{2,k}<c_k'\delta.
$$ 
We conclude that, if two connections $\n_1,\n_2\in \CO_{\omega,\ep}$ are gauge
equivalent, we can then choose $\delta>0$ to be sufficiently small that they are related by a gauge transformation in $\widetilde{\cG}_{\ep}$. Hence we deduce that the neighbourhoods
$\CO_{\omega,\ep}$ give local charts for $\mathcal{B}=\widetilde{\cC}_3/\widetilde{\cG}_4$.\endproof

\begin{prop}\label{pr:haus}
The manifold $\cB=\widetilde{\cC}_3/\widetilde{\cG}_4$ is Hausdorff.
\end{prop}

\proof 
Let $\nabla$, $\nabla_\omega$ be connections in $\widetilde\cC_3$ and $(\n_n)$, $(\n_n')$ sequences in
$\widetilde{\cC}_3$ such that: 
\begin{enumerate}[\hspace{0.5cm}(i)]
\item 
$\n_n\to \nom$ and $\n_n'\to \n'$ in the norm $\|\cdot\|_{2,3}$ as $n\to\infty$; 
\vspace{0.25cm}
\item 
for all $n$ there exist $U_n\in\cG_4$ such that $\n_n'=U_n\n_n U_n^*$.
\end{enumerate}
We show there exists $U \in \cG_4$ such that $U \nabla' U^* = \nabla$.
For appropriate elements $\alpha'$, $\alpha_n'$ and $\alpha_n$,
we write $\n'=\nom+\alpha'$, $\n_n'=\nom+\alpha_n'$ and $\n_n=\nom+\alpha_n$. 
By hypothesis $\alpha_n'\to\alpha'$ and $\alpha_n\to 0$ as $n\to\infty$
in the norm $\|\cdot\|_{2,3}$,
hence the sequences $(\|\alpha_n\|_{2,3})$ and $(\|\alpha_n'\|_{2,3})$ are uniformly bounded. 
From the proof of Theorem~\ref{pr:charts} it follows there exists a constant $C$ such that $\|U_n\|_{2,4}<C$ for all $n$, whence Theorem~\ref{th:techs}(iii)
implies that the sequence $(U_n)$ has a subsequence $(U_{n_m})$
converging to a point $U\in\Om^0_4(\End(\E))$ in the norm
$\|\cdot\|_{2,3}$. We compute that
\begin{align*}
\|[\nom,U_{n_m}]-[\nom,U_{n_r}]\|_{2,3}&=\|U_{n_m}\alpha_{n_m}-\alpha'_{n_m}U_{n_m}-U_{n_r}\alpha_{n_r}+\alpha'_{n_r}U_{n_r}\|_{2,3} \\
&\leq\|U_{n_m}\alpha_{n_m}-U_{n_r}\alpha_{n_r} + \alpha_{n_m}'  (U_{n_r} - U_{n_m})+(\alpha'_{n_r}-\alpha'_{n_m}) U_{n_r}\|_{2,3} \\
&\leq d_{4}C \left(\|\alpha_{n_m}\|_{2,3}+\|\alpha_{n_r}\|_{2,3}+\|\alpha'_{n_m}-\alpha'_{n_r}\|_{2,3}\right),
\end{align*}
where the last step uses the inequalities in Lemma~\ref{le:estim}. Each of the
terms on the right hand side can be made arbitrarily small, thus the sequence $([\nom,U_{n_m}])$ is Cauchy and hence convergent to $[\nom,U]$ in the norm $\|\cdot\|_{2,3}$. Combining this with convergence of $U_{n_m}$ to $U$ in $\|\cdot \|_{2,3}$-norm, it follows that $U_{n_m}\to U$ in the norm $\|\cdot\|_{2,4}$ and we conclude that
$U\in\cG_4$ and $U\n'U^*=\n$, so that finally the connections $\n'$ and $\nom$ are gauge equivalent.
\endproof

\section{Instantons on Toric Noncommutative Manifolds}\label{se:mod-sp}

In the previous section we studied the manifold structure of the space $\cB=\widetilde{\cC}_3/\widetilde{\cG}_4$. 
In this section we analyse the structure of the its subspace $\mM$ of self-dual $\U(2)$ connections on a four-dimensional toric noncommutative manifold $M_\theta$, modulo gauge transformations. 
Our strategy is to start with the infinitesimal geometry of $\mM$: 
by studying the linearised self-duality equations we obtain the
tangent space of the the moduli space at some base point $\nom$.
We then show that this infinitesimal structure can be integrated to describe the local structure of the moduli space near the base point $\nom$.

\subsection{Hodge structure on $M_\theta$} 
Of course, $M_\theta$ being a four-manifold means that
the differential graded algebra
$$
\Omega(M_\theta)=\oplus_{r=0}^\infty\Omega^r(M_\theta)
$$ 
is such that $\Omega^r(M_\theta)=0$ for $r>4$, with
$\Omega^4(M_\theta)$ one-dimensional. Moreover, the Hodge
operator $\star_\theta$ defined in Proposition~\ref{pr:hodge} maps two-forms to
two-forms,
$$
\star_\theta:\Omega^2(M_\theta)\to\Omega^2(M_\theta) .
$$
On such forms it obeys $\star_\theta^2=\id$ and so has eigenvalues $\pm 1$.

\begin{defn} On a toric noncommutative four-manifold $M_\theta$, a two-form
$\omega\in\Omega^2(M_\theta)$ is said to be {\em self-dual} if
$\star_\theta\omega=\omega$ or {\em anti-self-dual} if
$\star_\theta\omega=-\omega$. We write $\Om^2_+(M_\theta)$ or $\Om^2_-(M_\theta)$, respectively, for the spaces of self-dual forms and anti-self-dual forms, and 
we denote by $P_{\pm}$ the corresponding  projections
$$
P_{\pm}:\Om^2(M_\theta)\to\Om^2_{\pm}(M_\theta), \qquad P_{\pm} = \tfrac{1}{2} ( \id \pm \star_\theta) .
$$
\end{defn}

Let $\E$ be a Hermitian vector bundle over $M_\theta$ associated to a
principal $\U(2)$-bundle $P_\theta$ over $M_\theta$. 
Let $\n\in\mathcal{C}(\E)$ be
a compatible connection on $\E$. The curvature $F_\n:=\n^2$ of $\n$ is
an element of $\End^\textup{s}(\E\otimes_\A\Omega^2(M_\theta))\simeq
\Omega^2(M_\theta)\otimes_\A \End^{\textup{s}}(\E)$, {\em i.e.} a
two-form with values in the endomorphism bundle
$\textup{End}^\textup{s}(\E)$, leading to the following definition.

\begin{defn}\label{de:inst} An {\em instanton} on $\E$ is
a compatible connection $\n:\E\to\E\otimes_\A\Omega^1(M_\theta)$
whose curvature $F_\n=\n^2$ is a self-dual two-form
for the Hodge operator 
$$
\star_\theta \otimes \, \id
:\Omega^2(M_\theta)\otimes_\A\End^{\textup{s}}(\E)\to\Omega^2(M_\theta)\otimes_\A \End^{\textup{s}}(\E).
$$
\end{defn}

Given an irreducible compatible connection $\n\in \widetilde{\cC}$ with curvature $F_\n$, we write 
$[\n]$ for the equivalence class of $\n$ in the quotient space $\cB=\widetilde{\cC}/\widetilde{\cG}$.

\begin{defn} The {\em moduli space of instantons} on the vector
bundle $\E$ is the set
$$
\mM:=\left\{ [\n]\in\cB~|~F_\n~\text{is a self-dual two-form}\right\}.
$$
\end{defn}

\begin{rem}
\textup{It is quite possible that the moduli space $\mM$ on a given vector bundle $\E$ is in fact empty. In the remainder of the paper we shall assume that this is not the case.}
\end{rem}

A key property of the geometry of classical (spin) four-manifolds is their very special spin
structure. As we shall now see, the same is true for toric noncommutative four-manifolds. In the remainder of this section, we write $K:={\rm Spin}(4)\simeq\SU(2)\times\SU(2)$ and use the shorthand notation $K=K^+\times K^-$ with $K^\pm=\SU(2)$. We also write $V^\pm_j$ for the irreducible $K^\pm$-module with complex dimension $2j+1$, where $j=0, \frac{1}{2}, 1, \tfrac{3}{2}, \ldots$ .

\begin{prop}\label{pr:K-diffs}
Let $M_\theta$ be a four-dimensional toric noncommutative manifold. Then there exists a noncommutative principal bundle $\Cinf(M_\theta)\hookrightarrow \Cinf(P_\theta)$ with structure group $K$ such that, for each $r=0,1,2,\ldots$, there is an isomorphism of $\Cinf(M_\theta)$-bimodules, 
$$
\Om^r(M_\theta)\simeq \iGamma(P_\theta,\wedge^r(V_{\frac{1}{2}}^+ \otimes V_{\frac{1}{2}}^- )) .
$$
\end{prop}

\proof 
The required noncommutative principal bundle $P_\theta$ is the spin bundle $\Sigma_\theta$, so that the desired isomorphisms follow from the isomorphism given by the Clifford multiplication 
$$
\gamma_{\frac{1}{2}} : V_1 \to \textup{Hom}(V_{\frac{1}{2}} ^+, V_{\frac{1}{2}} ^-)
$$  
where, in the notation of \S\ref{sect:weitzenboeck}, we have $V_{\frac{1}{2}}  \simeq V_{\frac{1}{2}} ^+ \oplus V_{\frac{1}{2}} ^-$ and $V_1$ the typical fibre of the cotangent bundle. 
\endproof

Since $M$ is an even-dimensional manifold, the $\ZZ_2$-grading on the associated spinor bundle $\mathcal{S}$ yields a decomposition $\mathcal{S}=\mathcal{S}^+\oplus\mathcal{S}^-$ into even and odd spinors. The spaces of smooth sections of these bundles are in turn given by $\iGamma(M,\mathcal{S}^\pm)=\iGamma(P,V^\pm_{\frac{1}{2}} )$. 
Moreover, just as in the classical case, the decomposition of $K$-modules
$$
\wedge^2(V_{\frac{1}{2}}^+ \otimes V_{\frac{1}{2}}^- )=(V^+_1\otimes V^-_0)\oplus (V^+_0\otimes V^-_1)\simeq V^+_1\oplus V^-_1
$$
yields an isomorphism of $\Cinf(M_\theta)$-bimodules 
\begin{align}\label{two-dec}
\Om^2(M_\theta)&=\Cinf(P_\theta)\boxtimes_K(\wedge^2(V_{\frac{1}{2}}^+ \otimes V_{\frac{1}{2}}^- )) =\left(\Cinf(P_\theta)\boxtimes_{K^+} V^+_1\right) \oplus  \left(\Cinf(P_\theta)\boxtimes_{K^-} V^-_1\right)\\
\nonumber &=\Om^2_+(M_\theta)\oplus \Om^2_-(M_\theta).
\end{align}
In turn, this decomposition gives rise to the following useful expressions.

\begin{prop}\label{pr:spin-iso}
There are $\Cinf(M_\theta)$-bimodule isomorphisms
\begin{align*}
\Om^1(M_\theta)&\simeq\iGamma(M_\theta,\mathcal{S}_+\otimes\mathcal{S}_-)\simeq
\iGamma(M_\theta,\mathcal{S}_+)\otimes_{\A}
\iGamma(M_\theta,\mathcal{S}_-), \\
\Om^0(M_\theta)\oplus\Om^2_-(M_\theta)&\simeq
\iGamma(M_\theta,\mathcal{S}_-\otimes\mathcal{S}_-)\simeq\iGamma(M_\theta,\mathcal{S}_-)\otimes_{\A}\iGamma(M_\theta,\mathcal{S}_-).
\end{align*}
\end{prop}

\proof When $M$ carries an isometric action of the torus
$\mathbb{T}^N$, it is assumed that this action lifts to an
action of a double cover $\widetilde{\mathbb{T}}^N$ on the
spinor bundle $\mathcal{S}$ for which the $\ZZ_2$-grading
$\mathcal{S}=\mathcal{S}_+\oplus\mathcal{S}_-$ is equivariant.
Both isomorphisms now follow from Lemma~\ref{pr:tensiso}: the first is immediate from the proof of Proposition~\ref{pr:K-diffs}, whereas the second follows from the Clebsch-Gordan
decomposition $V_{\frac{1}{2}}^- \otimes V_{\frac{1}{2}}^- \simeq V^-_0\oplus
V^-_1$.\endproof

In \S\ref{se:eq-classes} we analysed the local structure of the
manifold $\cB=\widetilde{\cC}_3/\widetilde{\cG}_4$ showing that it is a Hausdorff Banach manifold
with local charts $\CO_{\omega,\ep}$ as in Theorem~\ref{pr:charts}. For an element
$\n=\nom+\alpha$ to be an instanton, its curvature
$F_\n=F_{\nom+\alpha}$ must satisfy the condition $P_-F_\n=0$,
where $P_-$ is the projection
onto the subspace of anti-self-dual two-forms. Explicitly, this means that
\begin{equation}\label{sd}
P_-\left( \nom(\alpha)+[\alpha,\alpha]\right)=0, \qquad \textup{for} \quad \alpha\in\CO_{\omega,\ep}.
\end{equation}
Moreover, there is the additional constraint that the
perturbation $\nom\mapsto\nom+\alpha$ should be orthogonal to all
gauge transformations, which translates into the condition
\begin{equation}\label{orth}
\nom^*\alpha=0,\qquad  \textup{for} \quad \alpha\in\CO_{\omega,\ep}.
\end{equation} 
The infinitesimal versions of these equations are given in the next proposition.

\begin{prop}\label{pr:lin-sd}
The linearisation of the Eqs.~\eqref{sd} and \eqref{orth} at the base point $\nom$ are 
\begin{equation}\label{lin-sd}
P_-\nom(\alpha)=0, \qquad \nom^*(\alpha)=0, \qquad  \textup{for} \quad \alpha\in\CO_{\omega,\ep} .
\end{equation}
\end{prop}

\proof It is a straightforward computation to take a
one-parameter family of instantons $\n_t=\nom+\alpha_t$ and
substitute it into Eq.~\eqref{sd}, then differentiate with
respect to $t$ and evaluate at $t=0$. This gives the first
equation as stated. The gauge orthogonality equation is
unchanged from Eq.~\eqref{orth} upon linearisation.\endproof

\subsection{The moduli space of instantons on $M_\theta$}
We are ready to study the manifold structure of the moduli space $\mM$. 
We begin with the linearised versions \eqref{lin-sd} of the self-duality
equations: this will provide us with a model for the tangent
space to the moduli space $\mM$. Next, we shall go on to
show that this infinitesimal model can be integrated to a local
coordinate patch $\mM\cap\CO_{\omega,\ep}$ for the moduli space
at the base point $\nom$.

We think of solutions of the linearised equations \eqref{lin-sd} as infinitesimal instantons modulo infinitesimal gauge transformations. 
The following proposition neatly encoded them as elements of the
cohomology of a complex defined by the base point
connection $\nom$.

\begin{prop}\label{pr:sd-comp}
Solutions of the linearised equations \eqref{lin-sd} are in a one-to-one correspondence with 
elements of the first cohomology $H^1$ of the complex of Hilbert spaces
\begin{equation}\label{sd-comp}
0\to \Om^0_4(\ad(P_\theta))\xrightarrow{\D_0} \Om^1_3(\ad(P_\theta))
\xrightarrow{\D_1} \Om^2_{-,2}(\ad(P_\theta))\to 0,
\end{equation}
where $\D_0:=\nom$ and $\D_1:=P_-\circ\nom$ are Fredholm operators.
\end{prop}

\proof The sequence \eqref{sd-comp} is indeed
a complex, since $F_\omega:=F_{\nom}$ is self-dual and so
$$
\D_1 \circ \D_0 = P_-\nom\circ\nom=P_-F_{\omega}=0.
$$
Moreover, both $\D_0$ and $\D_1$ are Fredholm by virtue of Lemma \ref{le:fred}.
The first cohomology of the
complex is precisely the vector space $H^1:=
\Ker\,\D_1/ \im\,\D_0 \simeq \Ker\,\D_0^* \cap\Ker\,\D_1$. 
\endproof

\begin{prop}
\label{prop:complex-index}
The cohomology groups $H^i$ for $i=0,1,2$, of the complex \eqref{sd-comp} are finite-dimensional. 
Moreover, the map 
\begin{equation}\label{F-sum}
 \D_0^*+\D_1:\Om^1_3(\ad(P_\theta))\to  \Om^0_4(\ad(P_\theta))\oplus \Om^2_{-,2}(\ad(P_\theta))
 \end{equation}
is a Fredholm operator whose index coincides with the alternating sum of the vector 
space dimensions $h^i={\rm dim}\,H^i$, $i=0,1,2$:
\begin{equation}\label{altsum}
\ind (\D_0^*+\D_1)=-h^0+h^1-h^2.
\end{equation}
\end{prop}
\proof
The cohomology $H^0$ is precisely the kernel of the map $\D_0=\nom$. Since we are only considering irreducible connections, Lemma~\ref{le:iso} tells us that $H^0$ is nothing other than the space $\textup{H}^0_{\textup{dR}}(M_\theta)$ of constant functions on $M_\theta$ ({\em cf}. Definition \ref{defn:deRham}). 
The fact that $\textup{H}^0_{\textup{dR}}(M_\theta)$ and $\textup{H}^0_{\textup{dR}}(M)$ are isomorphic as vector spaces implies that their dimension equals the number of connected components of $M$. 

For the cohomologies $H^1$ and $H^2$, note that the operator $\D_1$ is Fredholm by Lemma \ref{le:fred} so that $H^2 = \Ker \D_1^*$ is finite dimensional. This also implies that $\im \D_0 \subset \Ker \D_1$ is finite-dimensional, so that $H^1$ is also finite-dimensional. 
Hence, all cohomology groups of the complex \eqref{sd-comp} are finite-dimensional whose alternating sum of dimensions is precisely computed by the index of $\D_0^* + \D_1$.
\endproof

Most of all we are interested in the dimension of the moduli space, which is just the dimension $h^1$ of the infinitesimal moduli space. This we would derive immediately from the index of $\D_0^* + \D_1$ were we able to compute the dimensions $h^0$ and $h^2$. 

The proof above of Proposition~\ref{prop:complex-index} shows that $h^0= \textup{dim}\,\textup{H}^0_{\textup{dR}}(M_\theta)$. Classically, the vanishing of $h^2$ is deduced from a certain positive curvature assumption on $M$ \cite{AHS78}. We will follow this argument in the $\theta$-deformed case and determine $\ker \D_1^*$ by employing the Weitzenb\"ock formula derived in Theorem \ref{thm:weitzenbock}. 
Consider the operator 
$$
\D_0 + \D_1^* : \Om^0_4(\ad(P_\theta))\oplus \Om^2_{-,2}(\ad(P_\theta)) \to
\Om^1_3(\ad(P_\theta)),
$$
the formal adjoint of the operator in Eq.~\eqref{F-sum}. Using Proposition~\ref{pr:spin-iso}, the latter can be identified with the operator
$$
\mathcal D^-_\omega : \Gamma^\infty(M_\theta, \cS_- \otimes \cS_- \otimes (\ad P_\theta)) \to \Gamma^\infty(M_\theta, \cS_+ \otimes \cS_- \otimes (\ad P_\theta)).
$$
Taking the sum of $\mathcal D^-_\omega$ with its adjoint $\mathcal D^+_\omega: = (\mathcal D^-_\omega)^*$, we obtain the twisted Dirac operator
$$
\mathcal D_\omega : \Gamma^\infty(M_\theta, \cS \otimes \cS_- \otimes (\ad P_\theta)) \to \Gamma^\infty(M_\theta, \cS \otimes \cS_- \otimes (\ad P_\theta)).
$$
If $\mathcal D^-_\omega \psi =0$ then $\mathcal D_\omega^2 \psi = 0$, so that from Theorem \ref{thm:weitzenbock} we find that
\begin{equation}
\label{eq:positivity}
( \nabla_{\E \otimes \cS} \psi, \nabla_{\E \otimes \cS} \psi) + ( \psi ,  \gamma_\theta(F_{\nabla_{\E}} )\psi ) + (\psi ,\gamma_\theta(F_{\n_\cS}) \psi) = 0.
\end{equation}
Now recall that the connection $\nabla_\E$ has self-dual curvature, {\em i.e.} $F_{\nabla_{\E}} \in \Omega^2_+(\ad P_\theta)$. Moreover, as in Proposition~\ref{pr:spin-iso} we have
$$
\gamma_\theta: \Omega^2_+(M_\theta) \hookrightarrow \Gamma^\infty(M_\theta, \End(\cS_+)),
$$
whence the self-dual two-forms $\Omega^2_+(M_\theta)$ acts as zero on the odd spinors $\Gamma^\infty(M_\theta, \cS_-)$. 

Since the Levi-Civita connection is undeformed ({\em cf}. Lemma \ref{le:gr-inv}) it follows from the classical case \cite{AHS78} that, in its Clifford action on $\Om^2_{-,2}(\ad(P_\theta))$, the curvature $F_{\n_\cS}$ obeys
$$
\gamma_\theta(F_{\n_\cS}) = \tfrac13 R,
$$
where $R$ is the scalar curvature of the classical manifold $M$. If $R>0$ all terms in Eq.~\eqref{eq:positivity} are positive for $\psi \in \Om^2_{-,2}(\ad(P_\theta))$. It follows that if $\D_1^* \psi = 0$ then $\psi = 0$, from which we deduce that $h^2 = 0$ if the manifold $M$ has scalar curvature $R>0$.

As already explained, the alternating sum $-h^0 + h^1 - h^2$ coincides with the index of the operator 
$ \D_0^*+\D_1$ defined in Eq.~\eqref{F-sum}, or, equivalently, the index of the twisted Dirac operator $\mathcal{D}_\omega$. The following theorem shows how to evaulate this index.

\begin{thm}\label{th:index}
The index of the operator $\mathcal{D_\omega}$ is independent of $\omega$ and given by
$$
\ind\cD_\omega=\left\la [M_\theta], \mathrm{ch}(\mathrm{ad}(P_\theta))\cdot\mathrm{ch}(\mathcal{S}_-)\right\ra,
$$
where $[M_\theta]$ is the $\textup{K}$-homology class of the spectral triple $$
(\Cinf(M_\theta),L^2(M_\theta,\mathcal{S}),D),$$
the map $\mathrm{ch}$ is the Chern-Connes character on the $\textup{K}$-theory $\textup{K}(\Cinf(M_\theta))$ and $\la\cdot,\cdot\ra$ denotes the canonical pairing between $\textup{K}$-homology and cyclic cohomology. 
In particular, the value of the index of $\cD_\omega$ coincides with its classical analogue.
\end{thm}
\proof
We have already established in Proposition~\ref{prop:complex-index} that, for any $\omega \in \Omega^1_3(\ad P_\theta)$ $\cD_\omega$ is Fredholm . Also, we can write
$$
\cD_\omega = \cD_0 + \omega^* + P_- \omega,
$$
where $\cD_0$ is expressed in terms of the canonical (and undeformed) connection $\nabla_0$ on the bundle 
$\ad (P_\theta)$ and the Levi-Civita connection on the spinor bundle $\cS$. Now we can connect $\nabla_\omega$ to $\nabla_0$ by a continuous path, $\nabla_{x \omega}$ with $0 \leq x \leq 1$, of Fredholm operators. Indeed, 
for non-zero $x$ we have $x \omega \in \Omega^1_3(\ad P_\theta)$ if and only if $\omega \in \Omega^1_3(\ad P_\theta)$. By continuity of the Fredholm index we conclude that $\ind \cD_\omega = \ind \cD_0$. Since the latter operator coincides with its classical counterpart, as do the corresponding Sobolev spaces, we conclude that the index is independent of $\theta$. 

That the index can be expressed as a pairing between K-homology and cyclic cohomology is the content of the Connes--Moscovici index formula \cite{cm}, which we may apply after having established that $\cD_0$ coincides with the Dirac operator $\tilde p D \tilde p$ with coefficients in the vector bundle  $\mathcal{S}_-\otimes\ad(P_\theta)$. Here, $\tilde p$ is the projection for the corresponding finite projective module $\iGamma(M_\theta, \mathcal{S}_-\otimes\ad(P_\theta))$. In order to see this, we use the spin bundle isomorphisms of Proposition~\ref{pr:spin-iso}, under which the operator $\cD_0$ is identified with the twisted Dirac operator
\begin{equation}\label{tw-dirac2}
\tilde p D\tilde p:\iGamma(M_\theta,\mathcal{S}_+\otimes\mathcal{S}_-\otimes\ad(P_\theta))\to\iGamma(M_\theta,\mathcal{S}_-\otimes\mathcal{S}_-\otimes\ad(P_\theta)),
\end{equation}
acting on smooth sections. Then, upon noting that $\tilde p D \tilde p$ as a map on $L^2$-spaces has kernel and cokernel consisting of smooth elements, it follows that the $\ind \cD_0$ can be computed by the Connes--Moscovici index pairing \cite{cm} for the operator $\tilde p D \tilde p$. 
\endproof

Finally it remains to check that this infinitesimal description of the moduli space can be `integrated' to give a local version of its manifold structure.

\begin{thm}\label{th:mod-loc}
Suppose the scalar curvature of the manifold $M$ is positive so that $h^2=0$ in the complex \eqref{sd-comp}.
Let $\nom$ be an irreducible connection on the bundle $\E$ with self-dual curvature. Then there is a neighbourhood $\CO$ of the origin in the vector space $H^1$ such that $\mM\cap\CO_{\omega,\ep}$ is diffeomorphic to $\CO$.
\end{thm}

\proof First recall that we have an orthogonal decomposition 
\begin{equation}\label{orth2}
\Om^1_3(\ad(P_\theta))=\Ker \nom^*\oplus \im \nom .
\end{equation}
For brevity we write $V:=\Ker\,\nom^*$.
The fact that $h^2=0$ means that the restricted map $\D_1:V\to \Om^2_-(\ad(P_\theta))$ is surjective.
So there exists an inverse $\D_1^{-1}:\Om^2_-(\ad(P_\theta))\to V$ mapping
onto the orthogonal complement to $\Ker\,\D_1$ in $V$. Thus we can define the map
$$
F:V\to V,\qquad F(\alpha)=\alpha+\D_1^{-1}P_-([\alpha,\alpha]).
$$
Its derivative $\D F$ at the origin is the identity map and so, by the inverse function
theorem, there is a neighbourhood of the origin in $V$ on
which $F$ is invertible. By choosing $\ep>0$ sufficiently
small, one can arrange for $F$ to be invertible on the
coordinate chart $\CO_{\omega,\ep}$.

Next, take the subset $\mM\cap\CO_{\omega,\ep}$ of elements in
$\CO_{\omega,\ep}$ that satisfy the self-duality equations
\begin{equation}\label{sd2}
P_-(\nom(\alpha)+[\alpha,\alpha])=0, \qquad \nom^*(\alpha)=0.
\end{equation}
We claim that the image of $\mM\cap\CO_{\omega,\ep}$ under $F$
is a neighbourhood $\CO$ of the origin in $H^1$, from which it
will follow from invertibility of $F$ that
$\mM\cap\CO_{\omega,\ep}\simeq\CO$. To this end we observe that,
for $\alpha\in \mM\cap\CO_{\omega,\ep}$, we can write
$\alpha=F^{-1}(\beta)$ for some $\beta\in V$.  We get:
$$
P_-\nom(\beta)=\D_1(F(\alpha))=\D_1(\alpha)+P_-([\alpha,\alpha])=\nom(\alpha)+P_-([\alpha,\alpha])=0.
$$
Moreover, we have that 
$$
\nom^*(\beta)=\nom^*(F(\alpha))=\nom^*(\alpha)+\nom^*\D_1^{-1}P_-([\alpha,\alpha])=0.
$$
By the above construction of the map $\D_1^{-1}$, the element $\D_1^{-1}P_-([\alpha,\alpha])$ is necessarily an element of $V=\Ker\,\nom^*$, whence
it follows that $\nom^*(\beta)=0$. From these arguments we deduce
that $\beta\in H^1$. Since $F$ is in particular a linear map,
the image of $\mM\cap\CO_{\omega,\ep}$ is a neighbourhood of the
origin in $H^1$, as claimed.\endproof

In this way, the moduli space $\mM$ of self-dual connections inherits the manifold structure of the quotient space $\cB=\widetilde{\cC}_3/\widetilde{\cG}_4$. As a consequence, we arrive at the following theorem.

\begin{thm}
Let $M_\theta$ be a toric noncommutative four-manifold deforming a manifold $M$ with positive scalar curvature. If $\E=\iGamma(M_\theta,E)$ is a $\U(2)$ vector bundle over $M_\theta$ then the moduli space $\mM$ of irreducible instantons on $\E$ is either empty or a smooth Hausdorff manifold 
with local charts given by $\mM\cap\CO_{\omega,\ep}$ and of dimension
$$
\mathrm{dim}\,\mM=\left\la [M_\theta], \mathrm{ch}(\mathrm{ad}(P_\theta))\cdot\mathrm{ch}(\mathcal{S}_-)\right\ra +\textup{dim}\,\textup{H}^0_{\textup{dR}}(M_\theta) .
$$
\end{thm}

\proof This is a simple consequence of Theorems~\ref{th:index} and \ref{th:mod-loc}, combined with the
values computed above for of $h^0$ and $h^2$.\endproof

\subsection{The noncommutative sphere $S^4_\theta$} To end the paper, we return to the basic example of the toric noncommutative four-sphere $S^4_\theta$ discussed in \S\ref{se:qhopf}.

\begin{thm}
Let $\E$ be a torus-equivariant $\U(2)$ vector bundle over $S^4_\theta$ with second Chern number
${\rm ch}_2(\E)=k$.
Then the moduli space $\mM$  of irreducible instantons on $\E$ is a smooth Hausdorff manifold of dimension $8k-3$.
\end{thm}

\proof
First we note that, for each value $k\in \ZZ$ of the second Chern number, the corresponding moduli space is always non-empty. Indeed, non-empty families of instantons on $S^4_\theta$ for every value of $k\in \ZZ$ were explicitly constructed in \cite{brvs}. Since the scalar curvature of $S^4$ is positive for the round metric, we have that $h^2 = 0$ at all points of the moduli space. Thus the dimension of the moduli space of irreducible instantons on $\E$ is given in terms of the index computed in Theorem~\ref{th:index} with the value $h^0 = \dim {\rm H}_\dR^0(S^4) =1$. The relevant index is computed either classically or in the $\theta$-deformed case to be
$$
{\rm Index}\,\mathcal{D}_\omega= \langle [M_\theta] , \,\mathrm{ch}_0(\cS_-)\cdot \mathrm{ch}_2(\ad (P_\theta)) \rangle - 
\langle [M_\theta] , \,\mathrm{ch}_2(\cS_-) \cdot\mathrm{ch}_0(\ad (P_\theta)) \rangle = 2 (4 k) - 4.
$$
If in doubt, the required computations for $\theta \neq 0$ may be found in \cite{lvs:pfns}. Substituting this value for the index into the formula \eqref{altsum}, together with the values $h^0 = 1$ and $h^2=0$, yields $8k-3$ for the dimension of the moduli space. 
\endproof

\end{document}